\documentclass[12pt,french,english]{article}
\usepackage[T1]{fontenc}
\usepackage{amsmath}
\usepackage{color}
\usepackage{graphicx}
\usepackage{amssymb}

\makeatletter

 \usepackage{amsthm, amsmath, amsfonts, amssymb}
 \theoremstyle{plain}    
 \newtheorem{thm}{Theorem}

 \theoremstyle{remark}    
 \newtheorem*{acknowledgement*}{Acknowledgement} 
 \theoremstyle{plain}    
 \newtheorem{prop}[thm]{Proposition} 

 \theoremstyle{plain}    
 \newtheorem{cor}[thm]{Corollary} 
 \theoremstyle{plain}    
 \newtheorem{lem}{Lemma} 

\usepackage{a4wide}

\def\defi{\stackrel{\rm def}{=}}

\usepackage{color}
\usepackage{graphics}

\usepackage{babel}
\addto\extrasfrench{\providecommand{\fg}{\ifdim\lastskip>\z@\unskip\fi~\frqq}}
\makeatother
\begin{document}

\title{Prequantum chaos: Resonances of the prequantum cat map }

\author{Frédéric Faure\textit{}%
\thanks{Institut Fourier 100, rue des Maths, BP 74, 38402 St Martin d'Heres
\protect \\
email: frederic.faure@ujf-grenoble.fr \protect \\
http://www-fourier.ujf-grenoble.fr/\textasciitilde{}faure%
}}

\maketitle
\begin{abstract}
Prequantum dynamics has been introduced in the 70' by Kostant-Souriau-Kirillov
as an intermediate between classical and quantum dynamics. In common
with the classical dynamics, prequantum dynamics transports functions
on phase space, but add some phases which are important in quantum
interference effects. In the case of hyperbolic dynamical systems,
it is believed that the study of the prequantum dynamics will give
a better understanding of the quantum interference effects for large
time, and of their statistical properties. We consider a linear hyperbolic
map $M$ in $\mbox{SL}\left(2,\mathbb{Z}\right)$ which generates
a chaotic dynamics on the torus. This dynamics is lifted on a prequantum
fiber bundle. This gives a unitary prequantum (partially hyperbolic)
map. We calculate its resonances and show that they are related with
the quantum eigenvalues. A remarkable consequence is that quantum
dynamics emerges from long time behaviour of prequantum dynamics.
We present trace formulas, and discuss perspectives of this approach
in the non linear case.


\end{abstract}
\tableofcontents{}

\section{Introduction}

Quantum chaos is the study of wave dynamics (quantum dynamics) and
its spectral properties, in the limit of small wavelength, when in
this limit, the corresponding classical dynamical system is chaotic
\cite{gutzwiller}. This limit is also denoted by $\hbar\rightarrow0$,
and called the semi-classical limit. The usual mathematical models
to study quantum chaos are models of hyperbolic dynamics, because
there, the classical chaotic features are important and quite well
understood (mixing, exponential decay of correlations, central limit
Theorem for observables, etc...) \cite{brin-02,katok_hasselblatt}.
On the quantum side, the semi-classical formula like the Gutzwiller
trace formula (respect. the Van-Vleck formula) give descriptions of
the quantum spectrum (respect. the description of the wave evolution)
in the semi-classical limit, in terms of sum of complex amplitudes
along different classical trajectories. One important problem in quantum
chaos is that these semi-classical formula are mathematically proved
only for finite time (versus $\hbar\rightarrow0$), whereas some numerical
experiments suggest that they could be valid for much larger time,
like $t\simeq1/\hbar^{\alpha}$, $\alpha>0$ \cite{semi3}\cite{dittes_94}%
\footnote{In \cite{fred-trace-06}, we show the validity of semi-classical formula
for time large as $t\simeq C\log\left(1/\hbar\right)$, for any $C>0$,
for a quantized hyperbolic non linear map on the torus. %
}, and a lot of work in the physics literature of quantum chaos are
based on this last hypothesis (\cite{eck95} for example). The main
difficulty to prove this hypothesis, is related to the fact that the
number of classical trajectories which enter in the semi-classical
formula increases exponentially fast with time, like $e^{\lambda t}$,
where $\lambda$ is the Lyapounov exponent, and this makes difficult
to have a control of the error terms. 

For large time the structure described by the classical orbits in
phase space is far much finer than the Planck cells%
\footnote{Planck cells are the {}``best resolution{}`` of phase space made
by quantum mechanics at the scale $\hbar$. The limitation is due
to the uncertainty principle.%
} The validity of the semi-classical formula could be due to some average
effects in the sum of the huge number of complex amplitudes, at the
scale of the Planck cells. One goal is to justify and understand these
averaging process. 

It is known that classical hyperbolic dynamical systems have trace
formula which are exact, even in the \emph{non linear} cases, \cite{baladi_livre_00}
page 97, \cite{fred-RP-06}. These trace formula give the trace of
so-called regularized \emph{transfer operator}, in terms of periodic
orbits. The eigenvalues of the regularized transfer operator are called
Ruelle-Pollicott resonances and are useful to describe convergence
towards equilibrium and the decay of time-correlation functions in
hyperbolic dynamical systems. A remarkable result in this theory,
and which could be useful to exploit in quantum chaos, is the exactness
of these trace formula. As these formula involve a sum over classical
orbits, they can be interpreted as an averaging process over these
orbits. We hope to be able to extend this formalism of classical dynamical
systems to the semi-classical setting, in order to have a better control
on the averaging process between complexes amplitudes for large time,
and possibly to suggest an appropriate statistical approach for quantum
chaos.

To follow this program we have to find a classical transfer operator
whose trace formula is the semi-classical trace formula, and then
be able to compare (in operator norm) this transfer operator with
the quantum evolution operator, in order to prove the validity of
the semi-classical trace formula for the quantum dynamics. This paper
is a first step towards this objective. We propose here such an operator,
and perform its study for a particular hyperbolic dynamics, namely
a linear hyperbolic map on the torus. However the objective is not
yet reached because \emph{linear} hyperbolic maps are very particular
and semi-classical trace formula are already exact. The raised problematic
is therefore not fully present in this paper, but it is the main motivation
for this work, and we think that this analysis can be extended to
the non linear case and will then reveal its interest.

The transfer operator we propose is the prequantum evolution operator.
The prequantum dynamics is a very natural dynamics at the border between
classical and quantum dynamics. Similarly to the classical dynamics,
prequantum dynamics transports functions on phase space (more precisely
sections of a bundle), but introduces some complex phases which are
determined by the actions of the classical trajectories. These phases
are known to govern interferences phenomena which are characteristic
of wave dynamics and quantum dynamics. However, the difference with
quantum dynamics is that there is no uncertainty principle in prequantum
dynamics, and this simplifies its study in an essential way. The uncertainty
principle (which is mathematically introduced by the choice of a complex
polarization, or a complex structure on phase space, see Section \ref{sub:The-quantum-Hilbert}),
introduces a cut-off in phase space at the scale of the Planck constant
$\hbar$. One consequence of the absence of this cut-off in the prequantum
setting, is that prequantum formula are exact. Another consequence
is that the prequantum Hilbert space is much larger than the quantum
one, and the hyperbolicity hypothesis on the dynamics implies that
the prequantum wave functions escape towards finer and finer scales.
This escape of the prequantum wave function from macroscopic scales
towards microscopic scales for large time is described by a discrete
set of {}``prequantum resonances''. Another way to say it, is that
the prequantum resonances describe the time decay of correlations
between smooth prequantum functions. The biggest prequantum resonance(s)
(i.e. those with greatest modulus) dominate for long time and describe
the part of the prequantum wave functions which remain at the macroscopic
scale (i.e. at a scale larger than the Planck cells $\hbar$). We
therefore expect a general relation between these outer prequantum
resonances and the quantum eigenvalues which describe the quantum
wave evolution. 

The role of the prequantum dynamics and the corresponding trace formula
for quantum dynamics has already been suggested by many authors \cite{cvitanovic_chaos_93}\cite{dittes_94}\cite{tanner_00},
in particular V. Guillemin in \cite{guillemin_77} page 504. 

In this paper, starting from a linear hyperbolic map on the torus,
we show how to define the hyperbolic prequantum map on the torus,
and establish a relation between the discrete resonance spectrum of
the prequantum map and the discrete spectrum of the quantum map, see
Theorem \ref{thm:spectre} page \pageref{thm:spectre}. In the conclusion,
we discuss some perspectives.

\begin{acknowledgement*}
The author would like to thank the {}``Classical and Quantum resonances
team'' Nalini Anantharaman, Viviane Baladi, Yves Colin de Verdiere,
Luc Hillairet, Frédéric Naud, Stéphane Nonnenmacher and Dominique
Spehner, and also Laurent Charles, Jens Marklof, San Vu Ngoc, M. Pollicott
for discussions related to this work. We acknowledge a support by
{}``Agence Nationale de la Recherche'' under the grant ANR-05-JCJC-0107-01.
The author gratefully acknowledges Jens Marklof for a nice hospitality
at Bristol for a meeting on quantum chaos during June 2006. 
\end{acknowledgement*}

\section{\label{sec:Statement-of-the}Statement of the results}

In this Section we state the main result of this paper, and discuss
some consequences. In the next Sections, we will give precise definitions
and recall basics of the prequantum dynamics.

\subsection{Prequantum resonances and quantum eigenvalues}

Let $M:\mathbb{T}^{2}\rightarrow\mathbb{T}^{2}$ be a \textbf{hyperbolic
linear map} on $\mathbb{T}^{2}=\mathbb{R}^{2}/\mathbb{Z}^{2}$, i.e.
$M\in\mbox{SL}\left(2,\mathbb{Z}\right)$, $\textrm{Tr}M>2$. This
map is Anosov (uniformly hyperbolic), with strong chaotic properties,
such as ergodicity and mixing, see \cite{katok_hasselblatt} p. 154. 

The prequantum line bundle $L$ is a Hermitian complex line bundle
over $\mathbb{T}^{2}$, with constant curvature $\Theta=i2\pi N\omega$,
where $\omega=dq\wedge dp$ is the symplectic two form on $\mathbb{T}^{2}$
and $N\in\mathbb{N}^{*}$ is the Chern index of the line bundle. $N$
is related to $\hbar$ by $N=1/\left(2\pi\hbar\right)$. The \textbf{prequantum
Hilbert space} is the space $\mathcal{\tilde{H}}_{N}\defi L^{2}\left(L\right)$
of $L^{2}$ sections of $L$. Note that $\mathcal{\tilde{H}}_{N}$
is infinite dimensional. The prequantum dynamics is a lift of the
map $M$ on the bundle $L$ which preserves the connection. This prequantum
dynamics induces a transport of sections, and defines a unitary operator
acting in $\mathcal{\tilde{H}}_{N}$ called \textbf{pre-quantum map}
$\tilde{M}$. (In the following sections, this operator will be denoted
by $\tilde{M}_{N}$).

The \textbf{quantum Hilbert space} $\mathcal{H}_{N}$ is the space
of anti-holomorphic sections of $L$ (after the introduction of a
complex structure on $\mathbb{T}^{2}$). Contrary to the prequantum
case, $\mathcal{H}_{N}$ is finite dimensional, and $\mbox{dim}\mathcal{H}_{N}=N$
from the Riemann-Roch Theorem. The \textbf{quantum map} $\hat{M}$
is obtained by Weyl quantization of $M$. It is a unitary operator
acting in $\mathcal{H}_{N}$ \cite{berry-hannay-80}\cite{keating-91b}\cite{debievre-05}.
The \textbf{quantum spectrum} is the set of the eigenvalues of $\hat{M}$
denoted by $\exp\left(i\varphi_{k}\right)$, $k=1,\ldots,N$.

\paragraph{Classical resonances: }

\begin{figure}[htbp]
\begin{centering}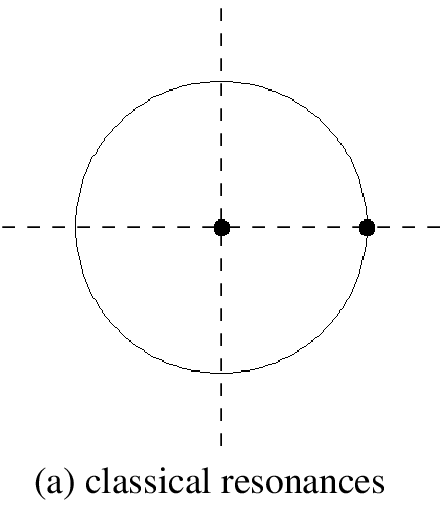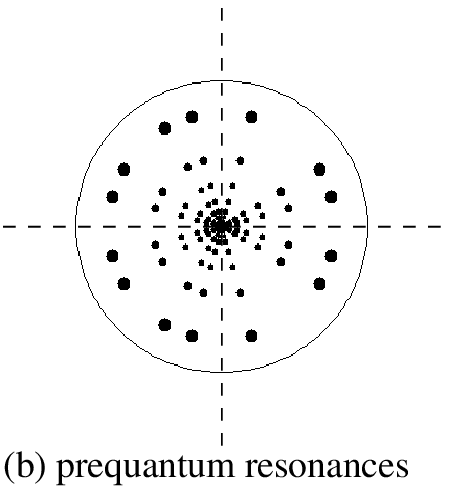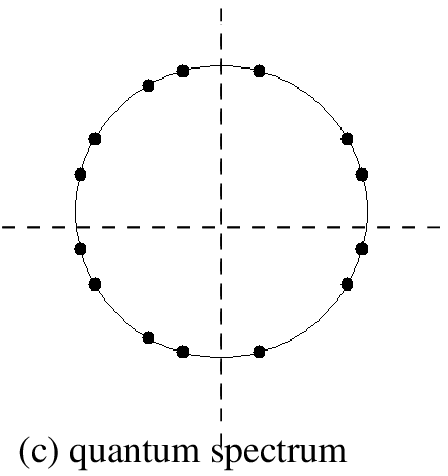\par\end{centering}

\caption{\label{fig:Comparison-of-spectra.}Comparison of spectra for the
linear cat map $M=\left(\protect\begin{array}{cc}
2 & 1\protect\\
1 & 1\protect\end{array}\right)$, with $N=1/\left(2\pi\hbar\right)=14$. }

(a) Ruelle-Pollicott resonances of the \textbf{classical map} $M$.
The isolated value $1$ traduces mixing property of the map. The absence
of resonances traduces super-exponential decay of time correlation
functions (See \cite{baladi_livre_00} page 225, or \cite{fred-RP-06}
for a simple description of the classical resonances as eigenvalues
of a trace class operator).

(b) Resonances $r_{n,k}$ of the \textbf{prequantum map} $\tilde{M}$,
calculated in this paper. $r_{n,k}=\exp\left(i\varphi_{k}-\lambda\left(n+1/2\right)\right)$,
$k=1\ldots N$, $n\in\mathbb{N}$. There are $N$ resonances on each
circle of radius $e^{-\lambda/2}e^{-\lambda n}$, $n\in\mathbb{N}$.

(c) Eigenvalues of the \textbf{quantum map} $\hat{M}$: $\exp\left(i\varphi_{k}\right)$,
$k=1\ldots N$.
\end{figure}

We first review the concept of time correlation functions and Ruelle
Pollicott resonances for the classical map $M$. These concepts give
a fruitful approach in order to study chaotic properties of the classical
dynamics, such as mixing or central limit theorem for observables,
etc... , see \cite{baladi_livre_00}. Let $\varphi,\phi\in\left(L^{2}\left(\mathbb{T}^{2}\right)\cap C^{\infty}\left(\mathbb{T}^{2}\right)\right)$,
and define the Transfer operator $M_{class.}$ acting on such functions
by $\left(M_{class}\varphi\right)\left(x\right)\defi\varphi\left(M^{-1}x\right)$,
$x\in\mathbb{T}^{2}$. For $t\in\mathbb{N}$, the \textbf{classical
time correlation function} is defined by:\[
C_{\phi,\varphi}\left(t\right)\defi\langle\phi|M_{class}^{t}\varphi\rangle\]

where the scalar product takes place in $L^{2}\left(\mathbb{T}^{2}\right)$.
Using the Fourier decomposition of $\phi,\varphi$, it is easy to
show that $C_{\phi,\varphi}\left(t\right)$ decreases with $t$ faster
than any exponential (see \cite{baladi_livre_00} p.226). I.e. for
any $\kappa>0$:\begin{equation}
C_{\phi,\varphi}\left(t\right)=\langle\phi|1\rangle\langle1|\varphi\rangle+o\left(e^{-\kappa t}\right)\label{eq:decay_class}\end{equation}
where $|1\rangle$ stands for the constant function $1$, and $\langle1|\varphi\rangle=\int_{\mathbb{T}^{2}}\varphi\left(x\right)dx$.
Eq.(\ref{eq:decay_class}) reveals the mixing property of the classical
map $M$. In order to study quantitatively the decay of $C_{\phi,\varphi}\left(t\right)$,
we introduce its Fourier transform:\[
\tilde{C}_{\phi,\varphi}\left(\omega\right)\defi\sum_{t\in\mathbb{N}}e^{it\omega}C_{\phi,\varphi}\left(t\right).\]
The \textbf{classical resonances of Ruelle-Pollicott} are $e^{i\omega}$
such that $\omega$ is a pole of the meromorphic extension of $\tilde{C}_{\phi,\varphi}\left(\omega\right)$,
$\omega\in\mathbb{C}$. They control the decay of $C_{\phi,\varphi}\left(t\right)$.
In our case, there is a simple pole $e^{i\omega}=1$, corresponding
to the mixing property, see figure \ref{fig:Comparison-of-spectra.}
(a). The super-exponential decay implies that there are non other
resonances. For a \emph{non linear} hyperbolic map we expect to observe
other resonances $e^{i\omega}$ , with $\left|e^{i\omega}\right|<1$,
see e.g. \cite{fred-RP-06}.

\paragraph{Prequantum resonances: }

We proceed similarly for the prequantum dynamics. Given two smooth
sections $\tilde{\varphi},\tilde{\phi}\in\left(L^{2}\left(L\right)\cap L^{\infty}\left(L\right)\right)$,
we define their \textbf{prequantum time-correlation function} by \[
C_{\tilde{\phi},\tilde{\varphi}}\left(t\right)\defi\langle\tilde{\phi}|\tilde{M}^{t}|\tilde{\varphi}\rangle,\qquad t\in\mathbb{N}\]
We wish to study the decay of $C_{\tilde{\phi},\tilde{\varphi}}\left(t\right)$.
The \textbf{prequantum resonances of Ruelle-Pollicott} are defined
as $e^{i\omega}$ such that $\omega$ is a pole of the meromorphic
extension of the Fourier transform of $C_{\tilde{\phi},\tilde{\varphi}}\left(t\right)$.
These resonances govern the decay of $C_{\tilde{\phi},\tilde{\varphi}}\left(t\right)$.
The main result of this paper is the following Theorem, illustrated
by figure \ref{fig:Comparison-of-spectra.}.

\vspace{0.5cm}\begin{center}\fbox{\parbox{15cm}{

\begin{thm}
\label{thm:spectre}Let $\tilde{M}$ be the prequantum map. There
exists an operator $\tilde{B}$, such that \[
\tilde{R}=\tilde{B}\tilde{M}\tilde{B}^{-1}\]
is defined on a dense domain of $L^{2}\left(L\right)$, and such that
$\tilde{R}$ extends uniquely to a Trace class operator in $L^{2}\left(L\right)$.
The eigenvalues of $\tilde{R}$, are the \textbf{prequantum resonances,}
and given by \begin{equation}
r_{n,k}=\exp\left(i\varphi_{k}-\lambda_{n}\right),\qquad k=1\ldots N,\quad n\in\mathbb{N}\label{eq:spectre_r_nk}\end{equation}
with $\exp\left(i\varphi_{k}\right)$ being the eigenvalues of the
quantum map $\hat{M}$ (\textbf{quantum eigenvalues)}, and $\lambda_{n}=\lambda\left(n+\frac{1}{2}\right)$,
with $\lambda$ being the Lyapounov exponent (i.e., $\exp\left(\pm\lambda\right)$
are the eigenvalues of $M$).
\end{thm}
}}\end{center}\vspace{0.5cm}

\paragraph{Remarks:}

\begin{itemize}
\item It is easy to see that the prequantum resonances are the eigenvalues
of $\tilde{R}$. Indeed, if $\tilde{\varphi},\tilde{\phi}\in\tilde{\mathcal{H}}_{N}=L^{2}\left(L\right)$
are sections which belong to the domains of $\tilde{B},\tilde{B}^{-1}$
respectively, then the time-correlation function $C_{\tilde{\phi},\tilde{\varphi}}\left(t\right)\defi\langle\tilde{\phi}|\tilde{M}^{t}|\tilde{\varphi}\rangle$,
$t\in\mathbb{N}$, can be expressed using the Trace class operator
$\tilde{R}$ as \[
C_{\tilde{\phi},\tilde{\varphi}}\left(t\right)\defi\langle\tilde{\phi}|\tilde{M}^{t}|\tilde{\varphi}\rangle=\left(\langle\tilde{\phi}|\tilde{B}^{-1}\right)\tilde{R}^{t}\left(\tilde{B}|\tilde{\varphi}\rangle\right).\]
Using a spectral decomposition of $\tilde{R}$, we deduce that the
discrete spectrum of $\tilde{R}$ gives the explicit exponential decay
of $C_{\tilde{\phi},\tilde{\varphi}}\left(t\right)$, and more precisely
that the eigenvalues of $\tilde{R}$ are the prequantum resonances
as defined above. 
\item The way we obtain the resonances of $\tilde{M}$ by conjugation with
a non unitary operator $\tilde{B}$ is well known in quantum mechanics
and is called the {}``complex scaling method'' \cite{simon_87}.
It is usually used in order to obtain the {}``quantum resonances
of open quantum systems''. Remind that $\tilde{M}$ is a unitary
operator. It will appear in the paper, that it has a continuous spectrum
on the unit circle. 
\end{itemize}

\paragraph{Sketch of the proof:}

The proof of Theorem \ref{thm:spectre} will be obtained in Section
\ref{sub:proof} page \pageref{sub:proof}. The main steps in the
proof is to show that the prequantum Hilbert space is unitary equivalent
to a tensor product $\mathcal{\tilde{H}}_{N}\equiv\mathcal{H}_{N}\otimes L^{2}\left(\mathbb{R}\right)$
involving the quantum Hilbert space $\mathcal{H}_{N}$ and a $L^{2}\left(\mathbb{R}\right)$
space (this is Eq. (\ref{eq:preq_space_torus}) page \pageref{eq:preq_space_torus}),
and then that the prequantum operator writes $\tilde{M}\equiv\hat{M}\otimes\exp\left(-i\hat{N}/\hbar\right)$,
where $\hat{M}$ is the quantum map acting in $\mathcal{H}_{N}$ and
$\hat{N}=Op_{Weyl}\left(\lambda qp\right)$ acting in $L^{2}\left(\mathbb{R}\right)$
is the Weyl quantization of a hyperbolic fixed point dynamics. It
is well known that $\exp\left(-i\hat{N}/\hbar\right)$ has a continuous
spectrum but a discrete set of resonances $\exp\left(-\lambda\left(n+1/2\right)\right)$,
$n\in\mathbb{N}$. So Eq.(\ref{eq:spectre_r_nk}) follows.

\subsection{Dynamical appearance of the quantum space}

For large time $t$, the $N$ external prequantum resonances on the
circle of radius $\exp\left(-\lambda/2\right)$ will dominate, and
with a suitable rescaling, $C_{\tilde{\phi},\tilde{\varphi}}\left(t\right)$
behaves for large time like quantum correlation functions, i.e. \textbf{matrix
elements of the quantum propagator}. More precisely:

\vspace{0.5cm}\begin{center}\fbox{\parbox{15cm}{

\begin{prop}
\label{pro:quantum_correl}if $\tilde{\phi},\tilde{\varphi}\in\tilde{\mathcal{H}}_{N}$
are prequantum wave functions, let us define $\phi=\hat{\Pi}\tilde{B}^{-1}\tilde{\phi}$,
$\varphi=\hat{\Pi}\tilde{B}\tilde{\varphi}$ , where $\hat{\Pi}=\tilde{\mathcal{H}}_{N}\rightarrow\mathcal{H}_{N}$
is the orthogonal projector called the Toeplitz projector (this requires
for $\tilde{\phi},\tilde{\varphi}$ to have suitable regularity so
that they belong to the corresponding domains). Then for large time
$t$ \[
\langle\tilde{\phi}|\tilde{M}^{t}|\tilde{\varphi}\rangle=\langle\phi|\hat{M}^{t}|\varphi\rangle e^{-\lambda t/2}\left(1+\mathcal{O}\left(e^{-\lambda t}\right)\right)\]
This means that \textbf{quantum dynamics emerges as the long time
behaviour of prequantum dynamics}.
\end{prop}
}}\end{center}\vspace{0.5cm}

The proof is given in Section \ref{sub:Relation-between-prequantum}
page \pageref{sub:Relation-between-prequantum}.

Let us make a comment on Theorem \ref{thm:spectre} and Proposition
\ref{pro:quantum_correl}. It is quite remarkable that the exterior
circle of prequantum resonances is identified with the quantum eigenvalues.
So the (generalized) eigenspace associated with these resonances is
equivalent to the quantum space. This unitary isomorphism appears
explicitely in the proof of the Theorem. In some sense, and this is
what Proposition \ref{pro:quantum_correl} shows, the quantum space
appears dynamically under the prequantum dynamics, and corresponds
to {}``long lived'' states. In this way the quantum dynamics appears
here without any quantization procedure, but by the prequantum dynamics
itself (which is itself a natural extension of the classical dynamics
as a lift on a line bundle).

\subsection{Trace formula}

As usual with Transfer operators, Trace formula express the trace
of a regularized transfer operator in terms of periodic orbits. The
prequantum unitary operator $\tilde{M}$ is not trace class, so the
Trace formula expresses the trace of $\tilde{R}^{t}$ which is trace
class. What is particular with the prequantum dynamics (compared to
classical dynamics), is the appearance of complex phases, related
with the classical actions of the periodic orbits.

\vspace{0.5cm}\begin{center}\fbox{\parbox{15cm}{

\begin{prop}
\label{pro:Trace_formula_Rt}For $t\in\mathbb{N}^{*}$, the trace
formula for the prequantum dynamics expresses the trace of $\tilde{R}^{t}$
in terms of periodic points of $M$ on $\mathbb{T}^{2}$ of period
$t$:\begin{equation}
\textrm{Tr}\left(\tilde{R}^{t}\right)=\sum_{x\equiv M^{t}x\,\left[\mathbb{Z}^{2}\right]}\frac{1}{\left|\textrm{det}\left(1-M^{t}\right)\right|}e^{iA_{x,t}/\hbar}\label{eq:Trace_M_preq}\end{equation}
where $A_{x,t}=\ointop\frac{1}{2}\left(qdp-pdq\right)+Hdt$ is the
classical action of the periodic orbit starting from $x=\left(q,p\right)$,
and $\left|\textrm{det}\left(1-M^{t}\right)\right|^{-1}=\left(e^{\lambda t/2}-e^{-\lambda t/2}\right)^{-2}$is
related with its instability. More explicitely, for a periodic point
characterized by $x=\left(q,p\right)\in\mathbb{R}^{2}$ and $M^{t}x=x+n$,
$n\in\mathbb{Z}^{2}$, then $A_{x,t}=\frac{1}{2}n\wedge x$.
\end{prop}
}}\end{center}\vspace{0.5cm}

The proof of Proposition \ref{pro:Trace_formula_Rt} is given in Section
\ref{sub:Proof-of-the_trace_formula}, and follows the usual procedure
to obtain trace formula for transfer operators (\cite{baladi_livre_00}
page 103 or \cite{fred-RP-06}). The idea is to use the fact that
the prequantum dynamics is a lift of the classical transport with
additional phases, and therefore use the Schwartz kernel of $\tilde{M}$.
Formally we write: \begin{equation}
\mbox{Tr}^{\flat}\left(\tilde{M}^{t}\right)=\int_{\mathbb{T}^{2}}dx\langle x|\tilde{M}^{t}|x\rangle=\int_{\mathbb{T}^{2}}dx\delta\left(M^{t}x-x\right)e^{iA_{x,t}/\hbar}=\sum_{x=M^{t}x}\frac{1}{\left|\textrm{det}\left(1-M^{t}\right)\right|}e^{iA_{x,t}/\hbar}\label{eq:calcul_trace}\end{equation}

This short calculation is made rigorous in the proof of Proposition
\ref{pro:Trace_formula_Rt} page \pageref{sub:Proof-of-the_trace_formula},
using a suitable regularization.

\paragraph{Relation with the quantum trace formula:}

\begin{cor}
From Eq.(\ref{eq:spectre_r_nk}), we deduce a relation between traces
of operators. For $t\in\mathbb{Z}$,\begin{equation}
\textrm{Tr}\left(\hat{M}^{t}\right)=\sqrt{\left|\textrm{det}\left(1-M^{t}\right)\right|}\textrm{Tr}\left(\tilde{R}^{t}\right)\label{eq:Traces}\end{equation}
and from Eq.(\ref{eq:Trace_M_preq}),
\end{cor}
\begin{equation}
\textrm{Tr}\left(\hat{M}^{t}\right)=\sum_{x=M^{t}x}\frac{1}{\sqrt{\left|\textrm{det}\left(1-M^{t}\right)\right|}}e^{iA_{x,t}/\hbar}\label{eq:Trace_M_q}\end{equation}

\begin{proof}
We have $\textrm{Tr}\left(\hat{M}^{t}\right)=\sum_{k=1}^{N}e^{i\varphi_{k}}$,
$\textrm{Tr}\left(\tilde{R}^{t}\right)=\sum_{k=1}^{N}\sum_{n\geq0}e^{i\varphi_{k}-\lambda_{n}}$
and $\sum_{n\geq0}e^{-\lambda_{n}t}=\sum_{n\geq0}e^{-\lambda\left(n+\frac{1}{2}\right)t}=\left(e^{\lambda t/2}-e^{-\lambda t/2}\right)^{-1}$,
and finally $\sqrt{\left|\textrm{det}\left(1-M^{t}\right)\right|}=\left(e^{\lambda t/2}-e^{-\lambda t/2}\right)$.
\end{proof}

\paragraph{Remarks:}

\begin{itemize}
\item Formula Eq.(\ref{eq:Trace_M_q}) can be proved directly, see e.g.
\cite{keating-91b}.
\item It is important to realize that the the trace formula for the quantum
operator Eq.(\ref{eq:Trace_M_q}) is exact in our case, because we
consider a \emph{linear} hyperbolic map $M$. For a non linear map
we expect that the trace formula for the prequantum map would be still
exact, whereas there is no more exact trace formula for the quantum
operator. What is known are semiclassical trace formula which give
$\mbox{Tr}\left(\hat{M}^{t}\right)$ in the limit $N\rightarrow\infty$,
but for relatively short time: $t=\mathcal{O}\left(\log N\right)$,
see \cite{fred-trace-06}. We give more comments on these trace formula
in the conclusion of this paper.
\end{itemize}

\section{Prequantum dynamics on $\mathbb{R}^{2}$\label{sec:Prequantum-dynamics-in}}

In this Section we recall the basics of prequantization on the euclidean
phase space $\mathbb{R}^{2}$. We will need this material in the next
Section. This is well known, see \cite{woodhouse2}, or \cite{borthwick_99}
for an introduction to geometric quantization on more general phase
spaces, i.e. Kähler manifolds.

\subsection{Hamiltonian dynamics}

We first start with a classical Hamiltonian flow. We consider the
phase space $\mathbb{R}^{2}$, and note $x=\left(q,p\right)\in\mathbb{R}^{2}$.
The symplectic two form is $\omega=dq\wedge dp$. A real valued \textbf{Hamiltonian
function} $H\in C^{\infty}\left(\mathbb{R}^{2}\right)$ defines a
Hamiltonian vector field $X_{H}$ by $\omega\left(X_{H},.\right)=dH$,
and given explicitly by\begin{equation}
X_{H}=\left(\frac{\partial H}{\partial p}\right)\frac{\partial}{\partial q}-\left(\frac{\partial H}{\partial q}\right)\frac{\partial}{\partial p}\label{eq:def_XH}\end{equation}
The Poisson bracket of $f,g\in C^{\infty}\left(\mathbb{R}^{2}\right)$
is $\left\{ f,g\right\} =\omega\left(X_{f},X_{g}\right)=X_{g}\left(f\right)=-X_{f}\left(g\right)$.
The vector field $X_{H}$ generates a Hamiltonian flow $\phi_{t}:\mathbb{R}^{2}\rightarrow\mathbb{R}^{2}$,
$t\in\mathbb{R}$. Explicitly, $\left(q\left(t\right),p\left(t\right)\right)=\phi_{t}\left(q\left(0\right),p\left(0\right)\right)$,
if $\frac{dq}{dt}=\frac{\partial H}{\partial p},\frac{dp}{dt}=-\frac{\partial H}{\partial q}$.
The flow transports functions: the action of $\phi_{t}$ on $f\in C^{\infty}\left(\mathbb{R}^{2}\right)$
is defined by $f_{t}\defi\left(f\circ\phi_{-t}\right)\in C^{\infty}\left(\mathbb{R}^{2}\right)$.
The corresponding evolution equation is $\frac{df_{t}}{dt}=\left\{ H,f_{t}\right\} =-X_{H}\left(f_{t}\right)$.
In order to explain the introduction of the prequantum operator below,
we rewrite this last equation as\begin{equation}
\frac{df_{t}}{dt}=-\frac{i}{\hbar}\left(-i\hbar X_{H}\right)f_{t}\label{eq:evolution_f_t}\end{equation}
where $\hbar>0$. A complex valued function $f\in C^{\infty}\left(\mathbb{R}^{2}\right)$
can be seen as a section of the trivial bundle $\mathbb{R}^{2}\times\mathbb{C}$
over $\mathbb{R}^{2}$. Prequantum dynamics we will define now, is
a generalisation of the transport of $f_{t}$ but for sections of
a non flat bundle over $\mathbb{R}^{2}$.

\subsection{The prequantum line bundle}

\begin{figure}[htbp]
\begin{centering}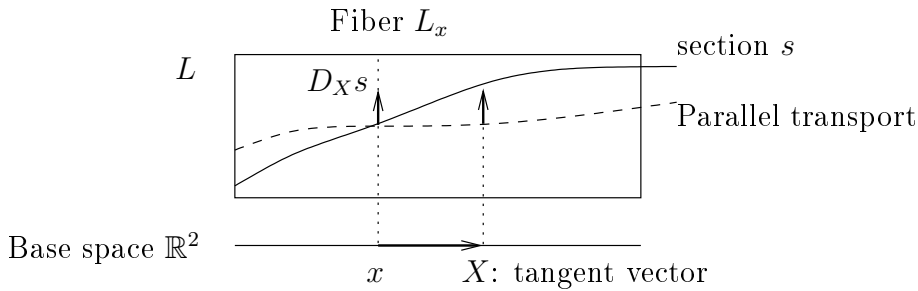\par\end{centering}

\caption{\label{fig:Covariant-derivative.}The covariant derivative of a section
$s$ with respect to a tangent vector $X$, is $D_{X}s\in L_{x}$
and characterizes the infinitesimal departure of the section $s$
from the parallel transport in the direction of $X$.}
\end{figure}

We introduce $\hbar>0$ called the {}``Planck constant'' and consider
a \textbf{Hermitian complex line bundle} $L$ over $\mathbb{R}^{2}$,
with a Hermitian connection%
\footnote{For a general introduction to Hermitian line bundles, see \cite{harris1}
p.71-77, or \cite{wells} p.67,p.77 %
} $D$. Each fiber $L_{x}$ over $x\in\mathbb{R}^{2}$ is isomorphic
to $\mathbb{C}$. A $C^{\infty}$ section $s$ of $L$ is a $C^{\infty}$
map $x\in\mathbb{R}^{2}\rightarrow s\left(x\right)\in L_{x}$. We
write $s\in A^{0}\left(L\right)$. The covariant derivative $D$ is
an operator $D:A^{0}\left(L\right)\rightarrow A^{1}\left(L\right)$
which acts on $C^{\infty}$ sections of $L$ and gives a $L$-valued
1-form. See figure \ref{fig:Covariant-derivative.}. We require that

\begin{figure}[htbp]
\begin{centering}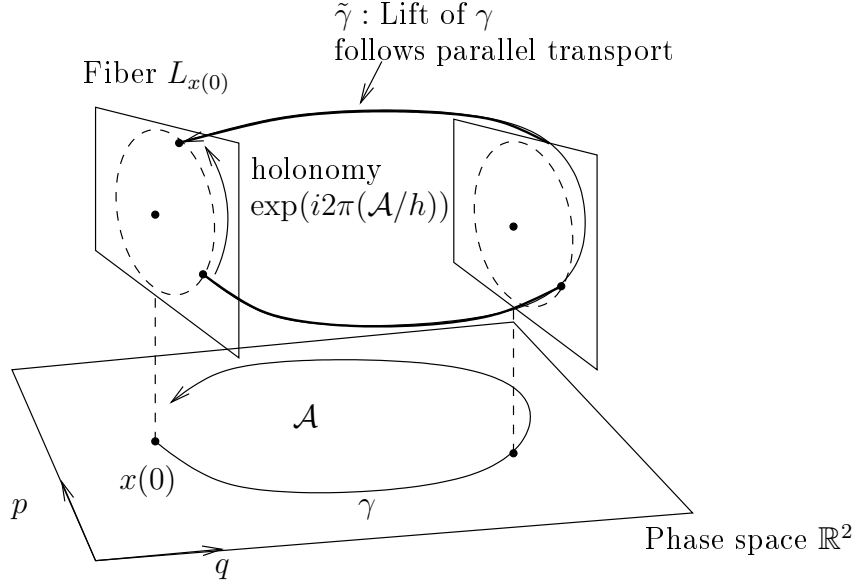\par\end{centering}

\caption{\label{cap:holonomy}A closed path $\gamma$ is lifted in the line
bundle following the parallel transport. The holonomy of the lifted
path $\tilde{\gamma}$ is equal to the phase $\exp\left(i2\pi\left(\mathcal{A}/h\right)\right)$
where $\mathcal{A}$ is the area of the closed path also called the
classical action of $\gamma$. $\left(\mathcal{A}/h\right)$ is called
the number of quanta enclosed in $\gamma$. These phases are responsible
for interference effects in quantum dynamics (wave dynamics).}
\end{figure}

\begin{enumerate}
\item \textbf{Leibniz's rule:} if $s\in A^{0}\left(L\right)$ is a section
of $L$, and $f\in C^{\infty}\left(\mathbb{R}^{2}\right)$ a function,
then $D\left(f.s\right)=df\otimes s+f.D\left(s\right)$.
\item If $h_{x}\left(.,.\right)$ denotes the Hermitian metric in the fiber
$L_{x}$, the connection $D$ should be compatible with $h$: $d\left(h\left(s_{1},s_{2}\right)\right)=h\left(Ds_{1},s_{2}\right)+h\left(s_{1},Ds_{2}\right)$.
In other words, if the section $s$ follows the connection in direction
$X$, i.e. $D_{X}s=0$, then $h\left(s,s\right)$ is constant in this
direction, i.e. $X\left(h(s,s)\right)=0$.
\item The curvature two form of the connection is\[
\Theta=\frac{i}{\hbar}\omega\]
where $\omega=dq\wedge dp$ is the symplectic two form. This last
condition means that the holonomy of a closed loop surrounding a surface
$\mathcal{S}\subset\mathbb{R}^{2}$ is $\exp\left(i\mathcal{\int_{\mathcal{S}}\omega}/\hbar\right)=\exp\left(i2\pi\left(\mathcal{A}/h\right)\right)$,
where $\left(\mathcal{A}/h\right)$ is interpreted as the number of
quanta $h=2\pi\hbar$ contained in the area $\mathcal{A}=\mathcal{\int_{\mathcal{S}}\omega}$.
See figure \ref{cap:holonomy}.
\end{enumerate}

\paragraph{A section of reference:}

As the base space $\mathbb{R}^{2}$ is contractible, we can choose
a unitary global section $r$ of $L$, i.e. such that $\left|r\left(x\right)\right|=\sqrt{h_{x}\left(r\left(x\right),r\left(x\right)\right)}=1$,
for every $x\in\mathbb{R}^{2}$. The section $r$ is called the \textbf{reference
section} and gives a trivialisation of the bundle $L$. We write its
covariant derivative $Dr=\theta r$, where $\theta$ is a 1-form on
$\mathbb{R}^{2}$. The requirements on $D$ above%
\footnote{The fact that $\theta$ is purely imaginary reflects the fact that
the connection is compatible with the Hermitian metric. Indeed, $h\left(r,r\right)=1$,
which gives $0=h\left(Dr,r\right)+h\left(r,Dr\right)\Leftrightarrow0=\mbox{Re}\left(h\left(r,\theta r\right)\right)=\mbox{Re}\left(\theta h\left(r,r\right)\right)=\mbox{Re}\left(\theta\right)$.
One requires that $\Theta=d\theta=i\omega/\hbar\Leftrightarrow d\eta=\omega$.%
} impose that $\theta=\frac{i}{\hbar}\eta$ with a real one form $\eta$
such that $d\eta=\omega$. In order to simplify some expressions below,
the section $r$ is chosen such that%
\footnote{The geometric meaning of $\eta$, also called the symmetric Gauge,
is that the reference section $r$ follows the parallel transport
along radial lines issued from the origin $x=0$. Indeed $\eta=\frac{1}{2}\left(qdp-pdq\right)\equiv\frac{1}{2}x\wedge dx$,
so if $X\in T_{x}\mathbb{R}^{2}\equiv\mathbb{R}^{2}$ is such that
$x\wedge X=0$, then $D_{X}r=\frac{i}{\hbar}\eta\left(X\right)r=0$. %
} \begin{equation}
\eta\defi\frac{1}{2}\left(qdp-pdq\right),\label{eq:def_eta}\end{equation}

With respect to the reference section $r$, any section $s\in A^{0}\left(L\right)$
is represented by a complex valued function $\psi$ on $\mathbb{R}^{2}$
defined by:\[
s\left(x\right)=\psi\left(x\right)r\left(x\right),\qquad\psi\left(x\right)\in\mathbb{C},\quad x\in\mathbb{R}^{2}\]

and $\left|s\left(x\right)\right|=\sqrt{h_{x}\left(s\left(x\right),s\left(x\right)\right)}=\left|\psi\left(x\right)\right|\sqrt{h_{x}\left(r\left(x\right),r\left(x\right)\right)}=\left|\psi\left(x\right)\right|$.

The space of interest for us, called the \textbf{prequantum Hilbert
space}, denoted by $L^{2}\left(L\right)$, is the space of sections
of $L$ with finite $L^{2}$ norm:

\vspace{0.5cm}\begin{center}\fbox{\parbox{15cm}{\begin{align}
L^{2}\left(L\right)\defi\left\{ s,\,\left\Vert s\right\Vert ^{2}=\int_{\mathbb{R}^{2}}dx\,\left|s\left(x\right)\right|^{2}<\infty\right\} \label{eq:Isom-s-Psi}\\
\cong L^{2}\left(\mathbb{R}^{2}\right)=\left\{ \psi,\,\int_{\mathbb{R}^{2}}dx\,\left|\psi\left(x\right)\right|^{2}<\infty\right\} , & \mbox{ with }s=\psi r\nonumber \end{align}

}}\end{center}\vspace{0.5cm}

where the last \emph{unitary isomorphism} is obtained by the identification
$s\equiv\psi$ given by Eq.(\ref{eq:Isom-s-Psi}). We will use this
unitary isomorphism all along the paper and work most of time with
the space $L^{2}\left(\mathbb{R}^{2}\right)$.

\paragraph{Remark:}

\begin{itemize}
\item If $\left\Vert s\right\Vert =1$, the function $\mbox{Hus}_{s}\left(x\right)=\left|s\left(x\right)\right|^{2}=\left|\psi\left(x\right)\right|^{2}$
is a probability measure on phase space $\mathbb{R}^{2}$ (i.e. $\int_{\mathbb{R}^{2}}\mbox{Hus}_{s}\left(x\right)dx=\left\Vert s\right\Vert ^{2}=1$),
and is called \textbf{Husimi distribution} of the section \textbf{$s$}
in the physic's literature \cite{hall99},\cite{folland-88}.
\end{itemize}

\subsection{The prequantum operator}

The \textbf{prequantum operator of Kostant-Souriau-Kirillov} acts
in the Hilbert space $L^{2}\left(L\right)$, Eq. (\ref{eq:Isom-s-Psi}),
and is defined by \begin{equation}
\mathbf{P}_{H}\defi-i\hbar D_{X_{H}}+H\label{eq:Def_P_H_phi_L}\end{equation}
where $D$ is the covariant derivative, $X_{H}$ is the Hamiltonian
vector field Eq.(\ref{eq:def_XH}), and $H$ denotes multiplication
of a section by the function $H$. If $H$ is a real function and
$X_{H}$ is complete, then $\mathbf{P}_{H}$ is a self-adjoint operator
(see \cite{woodhouse2}, page 162).

Writing $s=\psi r$ as in Eq.(\ref{eq:Isom-s-Psi}), we use Leibniz's
rule to write\begin{equation}
D_{X_{H}}\left(s\right)=D_{X_{H}}\left(\psi r\right)=d\psi\left(X_{H}\right)r+D_{X_{H}}\left(r\right)=\left(X_{H}\left(\psi\right)+\frac{i}{\hbar}\eta\left(X_{H}\right)\psi\right)r\label{eq:D_Xh}\end{equation}
and obtain that\[
\mathbf{P}_{H}\left(s\right)=\left(-i\hbar X_{H}\psi+\eta\left(X_{H}\right)\psi+H\psi\right)r=\left(P_{H}\psi\right)r\]
so $\mathbf{P}_{H}$ is isomorphic to the differential operator\begin{equation}
\boxed{P_{H}=-i\hbar X_{H}+\left(\eta\left(X_{H}\right)+H\right)}\label{eq:def_P_H_phi}\end{equation}
which acts in $L^{2}\left(\mathbb{R}^{2}\right)$. The last two terms
in Eq.(\ref{eq:Def_P_H_phi_L}) is the multiplication operator by
the function $\eta\left(X_{H}\right)+H=-\frac{1}{2}\left(q\left(\frac{\partial H}{\partial q}\right)+p\left(\frac{\partial H}{\partial p}\right)\right)+H$.
The role of the differential operator $P_{H}$ (respect. $\mathbf{P}_{H}$)
is to generate the {}``prequantum dynamics'', i.e. the evolution
of $\psi\left(t\right)\in L^{2}\left(\mathbb{R}^{2}\right)$ (respect.
$s\left(t\right)\in L^{2}\left(L\right)$) by the {}``\textbf{prequantum
Schrödinger equation}''\begin{equation}
\boxed{\frac{d\psi\left(t\right)}{dt}=-\frac{i}{\hbar}P_{H}\psi\left(t\right),\qquad\frac{ds\left(t\right)}{dt}=-\frac{i}{\hbar}\mathbf{P}_{H}s\left(t\right)}\label{eq:evolution_preq}\end{equation}
Whose solution is $\psi\left(t\right)=\tilde{U}_{t}\psi\left(0\right)$
(respect. $s\left(t\right)=\tilde{\mathbf{U}}_{t}s\left(0\right)$),
with the unitary operator in $L^{2}\left(\mathbb{R}^{2}\right)$:
\begin{equation}
\tilde{U}_{t}\defi\exp\left(-\frac{i}{\hbar}P_{H}t\right),\qquad\tilde{\mathbf{U}}_{t}\defi\exp\left(-\frac{i}{\hbar}\mathbf{P}_{H}t\right)\label{eq:U_t}\end{equation}
It can be shown that the term $H$ in Eq.(\ref{eq:Def_P_H_phi_L})
is necessary so that $\tilde{\mathbf{U}}_{t}$ preserves the connection
(see \cite{woodhouse2} page 163).

\paragraph{The Geometric and Dynamical phases:}

In this paragraph we interpret the terms which enter in the expression
of $P_{H}$, Eq.(\ref{eq:def_P_H_phi}). The reader can skip it and
go directly to Section \ref{sub:Canonical-basis-of}. According to
Eq.(\ref{eq:evolution_f_t}), the first term $\left(-i\hbar X_{H}\right)$
is just responsible for the transport of the function $\psi$ along
the Hamiltonian flow. The second term $\eta\left(X_{H}\right)$ comes
from the covariant derivative in Eq.(\ref{eq:D_Xh}), and without
the third term $H$, it would mean that the transported section $s\left(t\right)$
follows parallel transport over each trajectory $x\left(t\right)$.
The third term $H$ gives a \emph{departure from the parallel transport}.
The last two terms together change the value of the function $\psi\left(t\right)$
at point $x=\left(q,p\right)$ by the amount: \[
\left(\frac{d\psi}{dt}\right)_{(2)}\equiv\left(-\frac{i}{\hbar}\right)\left(\eta\left(X_{H}\right)+H\right)\psi\equiv\left(-\frac{i}{\hbar}\right)\left(\frac{1}{2}\left(q\frac{dp}{dt}-p\frac{dq}{dt}\right)+H\right)\psi\]
we recognize the infinitesimal action of the trajectory, see \cite{arnold-mmmc}.
As it is purely imaginary, it changes the phase of the function $\psi\left(t\right)$.
The first term related to the parallel transport over the trajectory
is called the {}``geometric phase'' in physics literature, whereas
the second term which depends explicitly on $H$ is called the {}``dynamical
phase''\cite{shapere}.

In order to be more precise, let $x\left(t\right)=\phi_{t}\left(x\left(0\right)\right)$,
$t\in\mathbb{R}$, be a trajectory on base space $\mathbb{R}^{2}$,
and $p\left(0\right)\in L_{x\left(0\right)}$ a point in the fiber
over the point $x\left(0\right)$. Let us denote $p_{\parallel}\left(t\right)$
the lifted path over $x\left(t\right)$ which starts from $p\left(0\right)$
and follows parallel transport. Then the prequantum dynamics is the
unique lifted path over $x\left(t\right)$ given by $p\left(t\right)=e^{\frac{-i}{\hbar}\int_{0}^{t}H\left(x\left(s\right)\right)ds}p_{\parallel}\left(t\right)$,
i.e. with a departure from the parallel transport given by the dynamical
phase. From that point of view, prequantum dynamics is a flow in the
fiber bundle $L$, which will be denoted by $p\left(t\right)=\tilde{\phi}_{t}p\left(0\right)$.
The unitary operator $\tilde{\mathbf{U}}_{t}$ defined in Eq.(\ref{eq:U_t}),
can be expressed by $\left(\tilde{\mathbf{U}}_{t}s\right)\left(x\left(t\right)\right)=\tilde{\phi}_{t}\left(s\left(x\left(0\right)\right)\right)$.

\begin{figure}[htbp]
\begin{centering}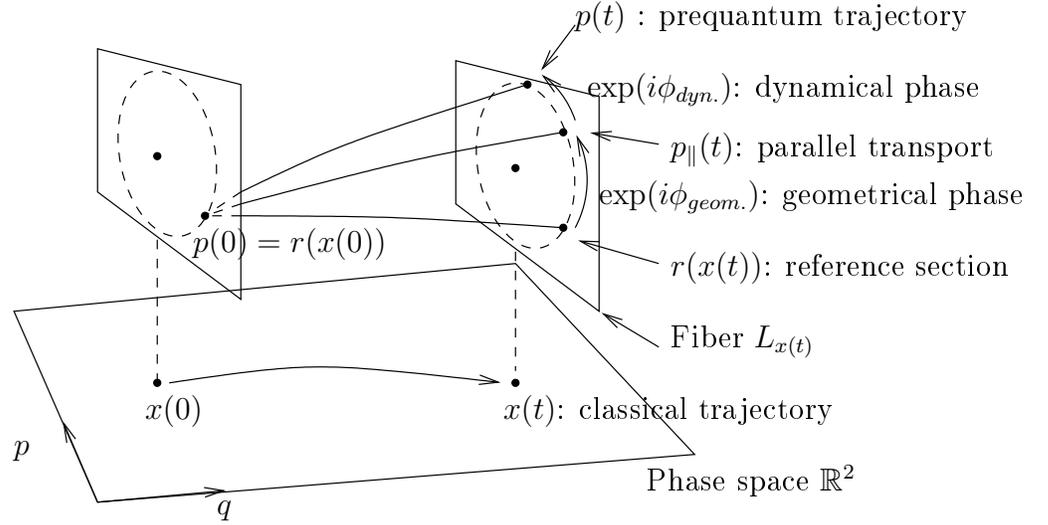\par\end{centering}

\caption{\label{cap:geom_phase}The prequantum dynamics is a lifted of the
classical dynamics $x\left(t\right)$ in phase space, where the lifted
path $p\left(t\right)$ follows the parallel transport $p_{\parallel}\left(t\right)$
with an additional phase $\phi_{dyn}=\frac{1}{\hbar}\int_{0}^{t}H\left(x\left(s\right)\right)ds$
called the dynamical phase. With respect to the reference section
$r\left(x\left(t\right)\right)$, the parallel transport is given
by $p_{\parallel}\left(x\left(t\right)\right)=e^{i\phi_{geom}}r\left(x\left(t\right)\right)$,
where $\phi_{geom}=-\frac{1}{2\hbar}\int\left(qdp-pdq\right)$ is
called the geometrical phase.}
\end{figure}

If $p\left(0\right)=r_{x\left(0\right)}$, then $p\left(t\right)$
is explicitely given with respect to the reference section $r_{x\left(t\right)}\in L_{x\left(t\right)}$
by:

\begin{equation}
p\left(t\right)=e^{-\frac{i}{\hbar}\int_{\gamma}dF}r_{x\left(t\right)}\label{eq:F_interpretation}\end{equation}
where $\gamma:x\left(0\right)\rightarrow x\left(t\right)$ is the
classical trajectory on the phase space $\mathbb{R}^{2}$ and $dF$
is the one-form on the extended phase space $\left(x,t\right)\in\mathbb{R}^{2}\times\mathbb{R}$:\begin{equation}
dF=\left(\eta\left(X_{H}\right)+H\right)dt=\frac{1}{2}\left(qdp-pdq\right)+Hdt\label{eq:dF}\end{equation}
which is the sum of the geometrical phase plus the dynamical phase.
See figure \ref{cap:geom_phase}.

In other terms, the solution $\psi\left(t\right)$ of Eq.(\ref{eq:evolution_preq}),
is given in terms of the classical flow by:\begin{equation}
\left(\tilde{U}_{t}\psi\right)\left(x\left(t\right)\right)=e^{-\frac{i}{\hbar}\int_{\gamma}dF}\psi\left(x\left(0\right)\right)\label{eq:Evolution_section}\end{equation}

\paragraph{Correspondence principle:}

An important interest for the prequantum operators comes from the
following proposition (see \cite{woodhouse2} page 157).

\begin{prop}
For any $f,g\in C^{\infty}\left(\mathbb{R}^{2}\right)$,\begin{equation}
[P_{f},P_{g}]=i\hbar P_{\{ f,g\}}\label{eq:Lie_algebra_homo}\end{equation}

\end{prop}
In other words, $f\in\left(C^{\infty}\left(\mathbb{R}^{2}\right),\left\{ ,\right\} \right)\rightarrow P_{f}\in\left(L\left(\mathcal{H}_{p}\right),\left[.,.\right]\right)$
is a \textbf{Lie algebra homomorphism}. In particular, it gives the
following basic commutation relation of quantum mechanics between
position and momentum, called the \textbf{{}``correspondence principle''}%
\footnote{Note that $P_{f=1}=\hat{\mbox{Id}}$ is obtained thanks to the third
term in (\ref{eq:def_P_H_phi}). $f\rightarrow\left(-X_{f}\right)$
is also a Lie algebra homomorphism (a more simple one), but $X_{f=1}=0\neq\hat{\mbox{Id}}$. %
}: $\left[P_{q},P_{p}\right]=i\hbar P_{\left\{ q,p\right\} }=i\hbar P_{1}=i\hbar\hat{\mbox{Id}}$. 

\begin{proof}
For any function $f,g\in C^{\infty}\left(M\right)$,$\left[X_{f},X_{g}\right]=-X_{\left\{ f,g\right\} }$.
If $\beta$ is a one form, and $X,Y$ two vector fields then $X\left(\beta\left(Y\right)\right)-Y\left(\beta\left(X\right)\right)=d\beta\left(X,Y\right)+\beta\left(\left[X,Y\right]\right)$
(see e.g. \cite{choquet-bruhat1} p. 207). With these two relations
we deduce:

\begin{eqnarray*}
\left[P_{f},P_{g}\right] & = & \left(-i\hbar\right)^{2}\left[X_{f},X_{g}\right]-i\hbar\left[X_{f},\eta\left(X_{g}\right)+g\right]-i\hbar\left[\eta\left(X_{f}\right)+f,X_{g}\right]\\
 & = & \hbar^{2}X_{\left\{ f,g\right\} }-i\hbar X_{f}\left(\eta\left(X_{g}\right)\right)+i\hbar\left\{ f,g\right\} +i\hbar X_{g}\left(\eta\left(X_{f}\right)\right)-i\hbar\left\{ g,f\right\} \\
 & = & \hbar^{2}X_{\left\{ f,g\right\} }+2i\hbar\left\{ f,g\right\} -i\hbar\left(d\eta\left(X_{f},X_{g}\right)+\eta\left(\left[X_{f},X_{g}\right]\right)\right)\\
 & = & \hbar^{2}X_{\left\{ f,g\right\} }+2i\hbar\left\{ f,g\right\} -i\hbar\omega\left(X_{f},X_{g}\right)-i\hbar\eta\left(\left[X_{f},X_{g}\right]\right)\\
 & = & i\hbar\left(-i\hbar X_{\left\{ f,g\right\} }+2\left\{ f,g\right\} -\left\{ f,g\right\} +\eta\left(X_{\left\{ f,g\right\} }\right)\right)=i\hbar P_{\left\{ f,g\right\} }\end{eqnarray*}

\end{proof}

\subsection{\label{sub:Canonical-basis-of}Canonical basis of operators in $L^{2}\left(\mathbb{R}^{2}\right)$}

In this Section we show that the Hilbert space $L^{2}\left(\mathbb{R}^{2}\right)$
(of prequantum sections, Eq.(\ref{eq:Isom-s-Psi})) is an irreducible
space for a convenient Weyl-Heisenberg algebra of operators constructed
with the covariant derivative. This will give a decomposition of the
space $L^{2}\left(\mathbb{R}^{2}\right)$ very useful for later use.

We have chosen coordinates $\left(q,p\right)\in\mathbb{R}^{2}$ on
phase space. Consider the covariant derivatives operators respectively
in the directions $\partial/\partial p$ and $\partial/\partial q$.
We denote them by:\[
\hat{Q}_{2}\defi-i\hbar D_{\frac{\partial}{\partial p}},\qquad\hat{P}_{2}\defi-i\hbar D_{\frac{\partial}{\partial q}}\]
With the unitary isomorphism Eq.(\ref{eq:Isom-s-Psi}), we identify
these operators with operators in $L^{2}\left(\mathbb{R}^{2}\right)$.
Using Eq.(\ref{eq:D_Xh}), and Eq.(\ref{eq:def_eta}), this gives
$\hat{Q}_{2}\equiv\left(-i\hbar\frac{\partial}{\partial p}+\eta\left(\frac{\partial}{\partial p}\right)\right)=\left(-i\hbar\frac{\partial}{\partial p}+\frac{1}{2}q\right)$.
Similarly for $\hat{P}_{2}$. We obtain:\begin{equation}
\boxed{\hat{Q}_{2}\equiv\left(-i\hbar\frac{\partial}{\partial p}\right)+\frac{1}{2}q,\qquad\hat{P}_{2}\equiv\left(-i\hbar\frac{\partial}{\partial q}\right)-\frac{1}{2}p.}\label{eq:def_Q2_P2}\end{equation}

Using the well known commutation relation $\left[q,\left(-i\hbar\frac{\partial}{\partial q}\right)\right]=i\hbar\hat{\mbox{Id}}$
(similarly with $p$), we deduce that $\left(\hat{Q}_{2},\hat{P}_{2},\hat{\mbox{Id}}\right)$
form a Weyl-Heisenberg algebra:\[
\left[\hat{Q}_{2},\hat{P}_{2}\right]=\hat{\mbox{Id}}.\]

In order to complete this algebra, define\begin{equation}
\hat{Q}_{1}\defi\mathbf{P}_{q},\qquad\hat{P}_{1}\defi\mathbf{P}_{p},\label{eq:def1_Q1_P1}\end{equation}
to be the prequantum operator for functions $q$ and $p$ respectively.
As before, the corresponding self-adjoint operators in $L^{2}\left(\mathbb{R}^{2}\right)$
are explicitely obtained from Eq.(\ref{eq:def_P_H_phi}):\begin{equation}
\boxed{\hat{Q}_{1}\equiv-\left(-i\hbar\frac{\partial}{\partial p}\right)+\frac{1}{2}q,\qquad\hat{P}_{1}\equiv\left(-i\hbar\frac{\partial}{\partial q}\right)+\frac{1}{2}p.}\label{eq:def_Q1_P1}\end{equation}
We directly check (or use Eq.(\ref{eq:Lie_algebra_homo})) that $\left[\hat{Q}_{1},\hat{P}_{1}\right]=i\hbar\hat{\mbox{Id}}$.
But also\[
\left[\hat{Q}_{i},\hat{P}_{j}\right]=i\hbar\hat{\mbox{Id}}\delta_{ij},\qquad\left[\hat{Q}_{i},\hat{Q}_{j}\right]=0,\quad\left[\hat{P}_{i},\hat{P}_{j}\right]=0.\]
So $\left(\hat{Q}_{1},\hat{P}_{1},\hat{Q}_{2},\hat{P}_{2},\hat{\mbox{Id}}\right)$
form a basis of the Weyl-Heisenberg algebra with {}``two-degree of
freedom'' in $L^{2}\left(\mathbb{R}^{2}\right)$. In fact, we have
obtained a new basis, from the original basis ($q,\left(-i\hbar\frac{\partial}{\partial q}\right)$,
$p,\left(-i\hbar\frac{\partial}{\partial p}\right),\hat{\mbox{Id}}$)
by a metaplectic transformation \cite{folland-88}. We summarize:

\begin{prop}
The space $L^{2}\left(\mathbb{R}^{2}\right)$ is an irreducible representation
space for the Weyl-Heisenberg algebra of operators $\left(\hat{Q}_{1},\hat{P}_{1},\hat{Q}_{2},\hat{P}_{2},\hat{\mbox{Id}}\right)$.
As a consequence we have a unitary isomorphism: \begin{equation}
\boxed{L^{2}\left(\mathbb{R}^{2}\right)\cong L^{2}\left(\mathbb{R}_{\left(1\right)}\right)\otimes L^{2}\left(\mathbb{R}_{\left(2\right)}\right)}\label{eq:H1_H2}\end{equation}
where $L^{2}\left(\mathbb{R}_{\left(1\right)}\right)$ (respect $L^{2}\left(\mathbb{R}_{\left(2\right)}\right)$)
denotes the Hilbert space of $L^{2}$ functions of one variable $f\left(Q_{1}\right),Q_{1}\in\mathbb{R}$
(respect. $f\left(Q_{2}\right),Q_{2}\in\mathbb{R}$), in which $\hat{Q}_{1}$
acts as $\left(\hat{Q}_{1}f\right)\left(Q_{1}\right)=Q_{1}f\left(Q_{1}\right)$
and $\left(\hat{P}_{1}f\right)\left(Q_{1}\right)=-i\hbar\frac{df}{dQ_{1}}\left(Q_{1}\right)$
(respect for $f\left(Q_{2}\right)$) . In other words, the decomposition
Eq.(\ref{eq:H1_H2}), means that $\psi\left(q,p\right)\in L^{2}\left(\mathbb{R}^{2}\right)$
is transformed into a function $\Psi\left(Q_{1},Q_{2}\right)\in L^{2}\left(\mathbb{R}_{\left(1\right)}\right)\otimes L^{2}\left(\mathbb{R}_{\left(2\right)}\right)$,
see Eq.(\ref{eq:formula_qp_Q1Q2}) below, for an explicit formula.
\end{prop}
We will see that the decomposition of the prequantum Hilbert space
Eq.(\ref{eq:H1_H2}) plays a major role for our understanding of the
prequantum dynamics.

\subsection{Case of a linear Hamiltonian function}

Consider the special case where $H$ is a linear function on $\mathbb{R}^{2}$,
with $v=\left(v_{q},v_{p}\right)\in\mathbb{R}^{2}$:\begin{equation}
H\left(q,p\right)=v_{q}p-v_{p}q\label{eq:H_linear}\end{equation}
then $X_{H}=v_{q}\frac{\partial}{\partial q}+v_{p}\frac{\partial}{\partial p}$
. The Hamiltonian flow after time 1 is a translation on $\mathbb{R}^{2}$
by the vector $v$, and we denote it by $T_{v}$: \begin{equation}
T_{v}\left(x\right)\defi x+v\label{eq:def_Tv_classique}\end{equation}

From definition Eq.(\ref{eq:def1_Q1_P1}) and linearity of Eq.(\ref{eq:Def_P_H_phi_L}),
we deduce that \begin{equation}
P_{H}=v_{q}\hat{P}_{1}-v_{p}\hat{Q}_{1}\label{eq:PH_translation}\end{equation}

The unitary operator generated by $P_{H}$ after time 1 will be written:
\begin{equation}
\tilde{T}_{v}\defi\exp\left(-\frac{i}{\hbar}P_{H}\right)=\exp\left(-\frac{i}{\hbar}\left(v_{q}\hat{P}_{1}-v_{p}\hat{Q}_{1}\right)\right)\label{eq:translation_Ph}\end{equation}
It is the prequantum lift of the classical translation Eq.(\ref{eq:def_Tv_classique}).

\paragraph{Remarks}

\begin{itemize}
\item The prequantum operator $P_{H}$, depends only on the operators $\hat{Q}_{1},\hat{P}_{1}$
and not on $\hat{Q}_{2},\hat{P}_{2}$. Therefore with respect to the
decomposition Eq.(\ref{eq:H1_H2}), operators $P_{H}$ and $\tilde{T}_{v}$
act trivially%
\footnote{We will see in Section \ref{sub:The-quantum-Hilbert}, that this is
related to the fact that translations on $\mathbb{R}^{2}$ preserve
the complex structure of $\mathbb{C}\equiv\mathbb{R}^{2}$.%
} in the space $L^{2}\left(\mathbb{R}_{\left(2\right)}\right)$, i.e.,
can be written as \begin{equation}
\tilde{T}_{v}=\tilde{T}_{v}^{\left(1\right)}\otimes\hat{\mbox{Id}}^{\left(2\right)}\label{eq:decomp_Tv}\end{equation}

\item The operator $P_{H}$ restricted to the space $L^{2}\left(\mathbb{R}_{\left(1\right)}\right)$
is identical to the Weyl-quantized operator $\mbox{Op}_{Weyl}\left(H\left(Q_{1},P_{1}\right)\right)=v_{q}\hat{P}_{1}-v_{p}\hat{Q}_{1}$,
see \cite{folland-88}.
\end{itemize}
\begin{prop}
The prequantum translation operators satisfy the algebraic relation
of the \textbf{Weyl-Heisenberg group}: for any $v,v'\in\mathbb{R}^{2}$,
\begin{equation}
\tilde{T}_{v}\,\tilde{T}_{v'}=e^{-iS/\hbar}\,\tilde{T}_{v+v'},\label{e:Heisenberg}\end{equation}
 with $S=\frac{1}{2}v\wedge v'=\frac{1}{2}\left(v_{1}v_{2}'-v_{2}v_{1}'\right)$
\end{prop}
\begin{proof}
There are two ways to see that. The first one (more algebraic) is
to use the explicit expression Eq.(\ref{eq:PH_translation}) of $P_{H}$
in terms of the operators $\hat{Q}_{1},\hat{P}_{1}$ and use $\left[\hat{Q_{1}},\hat{P_{1}}\right]=i\hbar\hat{Id}$
(Weyl-Heisenberg algebra) as well as the Baker-Campbell-Hausdorff
relation $e^{A}e^{B}=e^{A+B}e^{\frac{1}{2}\left[A,B\right]}$ for
any operators which satisfy $\left[A,B\right]=C.\hat{\mbox{Id}}$,
$C\in\mathbb{C}$. 

The second one (more geometrical) is to consider the initial point
$p=r_{x}\in L_{x}$ in the fiber over $x\in\mathbb{R}^{2}$. We want
to compute the phase $F$ obtained after a lift over the closed triangular
path $x\rightarrow\left(x+v'\right)\rightarrow\left(\left(x+v'\right)+v\right)\rightarrow x$
in the plane $\mathbb{R}^{2}$:\[
\tilde{T}_{v+v'}^{-1}\tilde{T}_{v}\tilde{T}_{v'}\left(p\right)=e^{-\frac{i}{\hbar}F}p\]
For a unique translation of $v$, starting at $x$, Eq.(\ref{eq:H_linear})
and Eq.(\ref{eq:dF}) give the phase \begin{equation}
F_{\mbox{trans.}}=\frac{1}{2}v\wedge x.\label{eq:F_translation}\end{equation}
So for the closed triangular path:\[
F=\frac{1}{2}\left(v'\wedge x\right)+\frac{1}{2}\left(v\wedge\left(x+v'\right)\right)+\frac{1}{2}\left(-\left(v+v'\right)\wedge\left(x+v+v'\right)\right)=\frac{1}{2}v\wedge v'\]

\end{proof}

\subsection{The quantum Hilbert space\label{sub:The-quantum-Hilbert}}

The usual Hilbert space of quantum mechanics which corresponds to
the phase space $\left(q,p\right)\in\mathbb{R}^{2}$, is the space
of functions $\psi\left(q\right)\in L^{2}\left(\mathbb{R}\right)$
\cite{landau-mq}. The prequantum Hilbert space $L^{2}\left(\mathbb{R}^{2}\right)$,
Eq.(\ref{eq:Isom-s-Psi}), is obviously too large. The usual procedure
to construct the quantum Hilbert space from the prequantum one in
geometric quantization is to add a complex structure on the phase
space $\mathbb{R}^{2}$, called a complex polarization, which induces
a holomorphic structure on the line bundle $L$, and then consider
the subspace of anti-holomorphic sections of $L$, (see \cite{woodhouse2},
\cite{borthwick_99}). We will show below that this indeed gives the
{}``standard'' Hilbert space of quantum wave functions $\psi\left(q\right)$.

We consider the canonical complex structure $J$ on phase space $\left(q,p\right)\in\mathbb{R}^{2}$
defined by $J\left(\frac{\partial}{\partial q}\right)=\frac{\partial}{\partial p}$
. Then $x=\left(q,p\right)\in\mathbb{R}^{2}$ is identified with $z\in\mathbb{C}$
by%
\footnote{The factor $1/\sqrt{2\hbar}$ is just a matter of choice.%
}:\begin{equation}
z=\frac{1}{\sqrt{2\hbar}}\left(q+ip\right)\label{eq:coord_z}\end{equation}
The \textbf{quantum Hilbert space} is defined to be the space of \textbf{anti-holomorphic
sections}: \begin{equation}
\mathcal{H}\defi\left\{ \textrm{section}\, s\in L^{2}\left(L\right)\,/\, D_{X^{+}}s=0,\textrm{ for all }X^{+}\in T^{1,0}\left(\mathbb{C}\right)\right\} \label{eq:def_H_quantum}\end{equation}
where the space of tangent vector of type $\left(1,0\right)$ (holomorphic
tangent vector) at point $x\in\mathbb{R}^{2}$ is spanned by $X^{+}=\frac{\partial}{\partial q}-i\frac{\partial}{\partial p}=\sqrt{\frac{2}{\hbar}}\frac{\partial}{\partial z}$.

\paragraph{Characterization of the quantum Hilbert space $\mathcal{H}$:}

Let us define the usual {}``annihilation'' and {}``creation''
operators $a_{2},a_{2}^{\dagger}$ by:\[
a_{2}\defi\frac{1}{\sqrt{2\hbar}}\left(\hat{Q}_{2}+i\hat{P}_{2}\right),\quad a_{2}^{\dagger}\defi\frac{1}{\sqrt{2\hbar}}\left(\hat{Q}_{2}-i\hat{P}_{2}\right)\]
The three operators $\left(a_{2},a_{2}^{\dagger},\hat{Id}\right)$,
with the relation$\left[a_{2},a_{2}^{\dagger}\right]=\hat{Id}$, form
a Cartan basis for the Weyl-Heisenberg algebra of operators acting
in the space $L^{2}\left(\mathbb{R}_{\left(2\right)}\right)$, which
enters in the decomposition Eq.(\ref{eq:H1_H2}). Note also that the
introduction of this basis of operators is natural after the choice
of the complex structure Eq.(\ref{eq:coord_z}). Similarly the operators
$\left(a_{1},a_{1}^{\dagger}\right)$ can be constructed with respect
to the space $L^{2}\left(\mathbb{R}_{\left(1\right)}\right)$, but
we will not need them. We recall that there is an orthonormal basis%
\footnote{This orthonormal basis has a nice physical meaning: for a free particle
in configuration space $\mathbb{R}^{2}$, with a constant magnetic
field $B=\left(2\pi\hbar\right)^{-1}\omega$, the Hamiltonian is $\hat{H}=\frac{1}{2}\left(-i\hbar\partial/\partial q-\frac{1}{2}p\right)^{2}+\frac{1}{2}\left(-i\hbar\partial/\partial p+\frac{1}{2}q\right)^{2}=\frac{1}{2}\hat{P}_{2}^{2}+\frac{1}{2}\hat{Q}_{2}^{2}=a_{2}^{\dagger}a_{2}+\frac{1}{2}$,
whose eigenspaces are $L^{2}\left(\mathbb{R}_{\left(1\right)}\right)\otimes\left(\mathbb{C}|n_{2}\rangle\right)$
and eigenvalues $n_{2}+\frac{1}{2}$ called Landau levels. %
} of $L^{2}\left(\mathbb{R}_{\left(2\right)}\right)$ related to the
{}``Harmonic Oscillator'', with vectors denoted by $|n_{2}\rangle\in L^{2}\left(\mathbb{R}_{\left(2\right)}\right)$,
$n_{2}\in\mathbb{N}$ and defined by \[
|0_{2}\rangle\in\mbox{Ker}\left(a_{2}\right)\quad\mbox{(one dimensional space)}\]
\begin{equation}
a_{2}|n_{2}\rangle=\sqrt{n_{2}}|n_{2}-1\rangle,\qquad a_{2}^{\dagger}|n_{2}\rangle=\sqrt{n_{2}+1}|n_{2}+1\rangle,\qquad n_{2}\in\mathbb{N}\label{eq:op_a_ap}\end{equation}
\[
\left(a_{2}^{\dagger}a_{2}\right)|n_{2}\rangle=n_{2}|n_{2}\rangle\]

\begin{prop}
With the unitary isomorphism Eq.(\ref{eq:Isom-s-Psi}), a section
$s\in\mathbf{\mathcal{H}}$ (Eq. (\ref{eq:def_H_quantum})) is identified
with a function $\psi\in L^{2}\left(\mathbb{R}^{2}\right)$ such that
$a_{2}\psi=0$, but also with the Bargmann space of anti-holomorphic
functions with weight $e^{-z\overline{z}/2}$ \cite{hall99}\cite{folland-88}:
\begin{align}
\mathcal{H}\cong & \left\{ \psi\in L^{2}\left(\mathbb{R}^{2}\right),\quad\psi\in\textrm{Ker}\left(a_{2}\right)\right\} \label{eq:H_quantum}\\
\cong & \left\{ \psi\in L^{2}\left(\mathbb{R}^{2}\right)\,/\,\psi\left(q,p\right)=e^{-z\overline{z}/2}\varphi\left(\overline{z}\right),\quad\varphi\left(\overline{z}\right)\mbox{ anti-holomorphic}\right\} :\quad\mbox{Bargmann space}\end{align}
With Eq.(\ref{eq:op_a_ap}) and the unitary isomorphism Eq.(\ref{eq:H1_H2}),
we get unitary isomorphisms%
\footnote{We can introduce an orthogonal projector in the prequantum space onto
the quantum space, called the \textbf{Toeplitz projector:} \[
\hat{\Pi}:\, L^{2}\left(L\right)\rightarrow\mathcal{H}.\]
With the identifications given by the unitary isomorphisms Eq.(\ref{eq:H1_H2})
and Eq.(\ref{eq:H_quantum_L2R1}), $\hat{\Pi}$ is the projector in
the space $L^{2}\left(\mathbb{R}_{\left(1\right)}\right)\otimes L^{2}\left(\mathbb{R}_{\left(2\right)}\right)$
onto the linear subspace $L^{2}\left(\mathbb{R}_{\left(1\right)}\right)\otimes\left(\mathbb{C}|0_{2}\rangle\right)$,
and can be written \begin{equation}
\hat{\Pi}\equiv\hat{\mbox{Id}}_{\left(1\right)}\otimes\left(|0_{2}\rangle\langle0_{2}|\right)\label{eq:Projecteur_Toeplitz}\end{equation}
This projector is used in geometric quantization to defined Toeplitz
quantization rules, see \cite{borthwick_99}.%
}:\begin{equation}
\boxed{\mathcal{H}\cong L^{2}\left(\mathbb{R}_{\left(1\right)}\right)\otimes\left(\mathbb{C}|0_{2}\rangle\right)\cong L^{2}\left(\mathbb{R}_{\left(1\right)}\right)}\label{eq:H_quantum_L2R1}\end{equation}

where $\left(\mathbb{C}|0_{2}\rangle\right)$ denotes the one dimensional
space $\textrm{Span}\left(|0_{2}\rangle\right)$, and the second isomorphism
is related to the choice of a vector%
\footnote{In geometrical terms, the complex structure $J$ is associated to
the one dimensional space $\mathbb{C}|0_{2}\rangle$. More generally,
the space of all possible homogeneous complex structures on $\mathbb{R}^{2}$
(which is the hyperbolic half plane $\mathbb{H}$) is identified with
the so called squeezed coherent states, which are the orbit of the
space $\left(\mathbb{C}|0_{2}\rangle\right)$ under the action of
the metaplectic group $\mbox{Mp}\left(2,\mathbb{R}\right)$ (generated
by quadratic functions of $\hat{Q}_{2},\hat{P}_{2}$). %
} $|0_{2}\rangle\in\mbox{Ker}\left(a_{2}\right)$.
\end{prop}
\begin{proof}
If $s=\psi r$, and $X^{+}=\frac{\partial}{\partial q}-i\frac{\partial}{\partial p}=\sqrt{\frac{2}{\hbar}}\frac{\partial}{\partial z}$,
then\[
-i\hbar D_{X^{+}}s=-i\hbar D_{\frac{\partial}{\partial q}}s-i\left(-i\hbar\right)D_{\frac{\partial}{\partial p}}s=\left(\left(\hat{P}_{2}-i\hat{Q}_{2}\right)\psi\right)r=-i\sqrt{2\hbar}\left(a_{2}\psi\right)r\]
so $D_{X^{+}}s=0\Leftrightarrow a_{2}\psi=0\Leftrightarrow\psi\in\mbox{Ker}\left(a_{2}\right)$.
Also we can write $D_{X^{+}}s=-i\sqrt{2\hbar}\left(\frac{\partial\psi}{\partial z}+\frac{1}{2}\overline{z}\psi\right)r$,
and $D_{X^{+}}s=0\Leftrightarrow\frac{\partial\psi}{\partial z}=-\frac{1}{2}\overline{z}\psi\Leftrightarrow\psi=e^{-z\overline{z}/2}\varphi\left(\overline{z}\right)$,
with an anti-holomorphic function $\varphi\left(\overline{z}\right)$.
\end{proof}

\paragraph{Correspondence with the usual Quantum Hilbert space $L^{2}\left(\mathbb{R}\right)$:}

We can make the connection between the space $\mathcal{H}$ and the
usual space of quantum wave functions more explicit. In {}``standard
quantum mechanics'' also called {}``position representation'',
the quantum Hilbert space associated to the phase space $\left(q,p\right)\in\mathbb{R}^{2}\equiv T^{*}\mathbb{R}$
consists of wave functions $\varphi\left(q\right)\in L^{2}\left(\mathbb{R}\right)$.
In this Section, we show that this space $L^{2}\left(\mathbb{R}\right)$
coincides with the space $L^{2}\left(\mathbb{R}_{\left(1\right)}\right)$
used in Eq.(\ref{eq:H_quantum_L2R1}). For that purpose we have to
show that the map $\varphi\in L^{2}\left(\mathbb{R}_{\left(1\right)}\right)\rightarrow\psi\in\mathcal{H}\subset L^{2}\left(\mathbb{R}^{2}\right)$
coincides with the \textbf{Bargmann Transform} \cite{hall99} of $\varphi$. 

\begin{prop}
If $\varphi\in L^{2}\left(\mathbb{R}_{\left(1\right)}\right)$, the
isomorphism $\mathcal{H}\cong L^{2}\left(\mathbb{R}_{\left(1\right)}\right)$
in Eq.(\ref{eq:H_quantum_L2R1}) is given by $\varphi\in L^{2}\left(\mathbb{R}_{\left(1\right)}\right)\rightarrow\psi\in\mathcal{H}\subset L^{2}\left(\mathbb{R}^{2}\right)$,
with\[
\psi\left(q,p\right)=\frac{1}{\left(\pi\hbar\right)^{1/4}}e^{iqp/\left(2\hbar\right)}\int_{\mathbb{R}}dQ_{1}\varphi\left(Q_{1}\right)e^{-iQ_{1}p/\hbar}e^{-\left(Q_{1}-q\right)^{2}/\left(2\hbar\right)}\]
We recognize the \textbf{Bargmann transform} \cite{hall99} of $\varphi$.
\end{prop}
\begin{proof}
From Eq.(\ref{eq:def_Q1_P1}),Eq.(\ref{eq:def_Q2_P2}), we have an
explicit relation between the representation of a function $\psi$
in $\left(q,p\right)$ variables or $\left(Q_{1},Q_{2}\right)$ variables:
\begin{equation}
\psi\left(q,p\right)=\int dQ_{1}dQ_{2}\langle qp|Q_{1}Q_{2}\rangle\Psi\left(Q_{1},Q_{2}\right)\label{eq:formula_qp_Q1Q2}\end{equation}
with\[
\langle qp|Q_{1}Q_{2}\rangle\defi\delta\left(Q_{1}+Q_{2}-q\right)e^{i\frac{1}{2}\left(Q_{2}-Q_{1}\right)p/\hbar}\]
(which comes from $\langle p_{0}|\xi_{p}\rangle=e^{i\xi_{p}p_{0}/\hbar}$,
$\langle q_{0}|q\rangle=\delta\left(q_{0}-q\right)$ and $q=Q_{1}+Q_{2}$,
$\xi_{p}=\frac{1}{2}\left(Q_{2}-Q_{1}\right)$). Now if $\psi\in\mathcal{H}$,
then from Eq.(\ref{eq:H_quantum_L2R1}), $\Psi\left(Q_{1},Q_{2}\right)=\varphi\left(Q_{1}\right)\varphi_{0}\left(Q_{2}\right)$,
where $\varphi_{0}\left(Q_{2}\right)=\langle Q_{2}|0_{2}\rangle=\left(\pi\hbar\right)^{-1/4}\exp\left(-Q_{2}^{2}/\left(2\hbar\right)\right)$.
This gives\begin{eqnarray*}
\psi\left(q,p\right) & = & \int dQ_{1}dQ_{2}\delta\left(Q_{1}+Q_{2}-q\right)e^{i\frac{1}{2}\left(Q_{2}-Q_{1}\right)p/\hbar}\varphi\left(Q_{1}\right)\frac{1}{\left(\pi\hbar\right)^{1/4}}\exp\left(-\frac{Q_{2}^{2}}{2\hbar}\right)\\
 & = & \frac{1}{\left(\pi\hbar\right)^{1/4}}e^{iqp/\left(2\hbar\right)}\int dQ_{1}e^{-iQ_{1}p/\hbar}\varphi\left(Q_{1}\right)\exp\left(-\frac{\left(q-Q_{1}\right)^{2}}{2\hbar}\right)\end{eqnarray*}

\end{proof}

\subsection{\label{sub:Case-of-a_Quadratic_H}Case of a quadratic Hamiltonian
function}

We consider now the special case where the Hamiltonian $H\left(q,p\right)$
is a quadratic function: \begin{equation}
H\left(q,p\right)=\frac{1}{2}\alpha q^{2}+\frac{1}{2}\beta p^{2}+\gamma qp,\qquad\alpha,\beta,\gamma\in\mathbb{R}\label{eq:H_quad}\end{equation}
Let us denote by $M\in\mbox{SL}\left(2,\mathbb{R}\right)$ the flow
on $\mathbb{R}^{2}$ generated by the quadratic Hamiltonian $H$ after
time 1 ($M$ is a linear symplectic map).

\begin{prop}
\label{pro:The-prequantum-operator_Quad}With respect to the decomposition
Eq.(\ref{eq:H1_H2}), the prequantum operator writes\begin{equation}
P_{H}=P_{H}^{\left(1\right)}\otimes Id_{\left(2\right)}+Id_{\left(1\right)}\otimes P_{H}^{\left(2\right)}\label{eq:PH_quadratic}\end{equation}
with\begin{equation}
P_{H}^{\left(1\right)}\defi\frac{1}{2}\alpha\hat{Q}_{1}^{2}+\frac{1}{2}\beta\hat{P}_{1}^{2}+\gamma\left(\frac{1}{2}\hat{Q}_{1}\hat{P}_{1}+\frac{1}{2}\hat{P}_{1}\hat{Q}_{1}\right)=Op_{Weyl}^{\left(1\right)}\left(H\right)\label{eq:PH_1_quad}\end{equation}
which acts in $L^{2}\left(\mathbb{R}_{\left(1\right)}\right)$, and\[
P_{H}^{\left(2\right)}\defi-\frac{1}{2}\alpha\hat{Q}_{2}^{2}-\frac{1}{2}\beta\hat{P}_{2}^{2}+\gamma\left(\frac{1}{2}\hat{Q}_{2}\hat{P}_{2}+\frac{1}{2}\hat{P}_{2}\hat{Q}_{2}\right)=Op_{Weyl}^{\left(2\right)}\left(H_{\left(2\right)}\right)\]
which acts in $L^{2}\left(\mathbb{R}_{\left(2\right)}\right)$. Here,
$Op_{Weyl}^{\left(i\right)}$ , $i=1,2$, means usual Weyl (symmetric)
quantization of quadratic symbols, with respectively $\left(Q_{1},P_{1}\right)$
or $\left(Q_{2},P_{2}\right)$. The function \begin{equation}
H_{\left(2\right)}\left(q,p\right)\defi-\frac{1}{2}\alpha q^{2}-\frac{1}{2}\beta p^{2}+\gamma qp\label{eq:H_2_quadrat}\end{equation}
 can be written as $H_{\left(2\right)}=-H\circ\mathcal{T}$ where
$\mathcal{T}\left(q,p\right)=\left(q,-p\right)$ is the {}``time
reversal'' operation.
\end{prop}
\begin{proof}
The Hamiltonian vector field is $X_{H}=\left(\gamma q+\beta p\right)\frac{\partial}{\partial q}-\left(\alpha q+\gamma p\right)\frac{\partial}{\partial p}$.
We compute then \begin{equation}
\eta\left(X_{H}\right)+H=0\label{eq:F_quadratic}\end{equation}
 so $P_{H}=-i\hbar X_{H}=\left(\gamma q+\beta p\right)\left(-i\hbar\frac{\partial}{\partial q}\right)-\left(\alpha q+\gamma p\right)\left(-i\hbar\frac{\partial}{\partial p}\right)$.
Note that this means that the prequantum transport by $P_{H}$ is
equivalent to the Hamiltonian transport Eq.(\ref{eq:evolution_f_t}).
Using Eq.(\ref{eq:def_Q2_P2}) and Eq.(\ref{eq:def_Q1_P1}), we deduce
the expression of $P_{H}$ in terms of the operators $\left(\hat{Q}_{i},\hat{P}_{i}\right)$.
\end{proof}

\paragraph{Remarks}

\begin{itemize}
\item The separation of terms in Eq.(\ref{eq:PH_quadratic}), has the following
direct consequence on the prequantum dynamics. Let \[
\tilde{M}_{\left(1\right),t}\defi\exp\left(-\frac{i}{\hbar}P_{H}^{\left(1\right)}t\right),\qquad\tilde{M}_{\left(2\right),t}\defi\exp\left(-\frac{i}{\hbar}P_{H}^{\left(2\right)}t\right)\]
be the unitary operators acting in $L^{2}\left(\mathbb{R}_{\left(1\right)}\right)$
and $L^{2}\left(\mathbb{R}_{\left(2\right)}\right)$ respectively,
and generated by $P_{H}^{\left(1\right)}$ and $P_{H}^{\left(2\right)}$
respectively. Then the total unitary operator in $L^{2}\left(\mathbb{R}^{2}\right)$
(the prequantum propagator) decomposes as a tensor product: \begin{equation}
\boxed{\tilde{M}_{t}\defi\exp\left(-\frac{i}{\hbar}P_{H}t\right)=\tilde{M}_{\left(1\right),t}\otimes\tilde{M}_{\left(2\right),t}}\label{eq:Mt_M1_M2}\end{equation}
We will see that this tensor product is the main phenomenon which
explains that the spectrum of prequantum resonances is a product of
two spectra in Eq.(\ref{eq:spectre_r_nk}).
\item Note that the prequantum evolution does not preserve the quantum Hilbert
space $\mathcal{H}\cong L^{2}\left(\mathbb{R}_{\left(1\right)}\right)\otimes\left(\mathbb{C}|0_{2}\rangle\right)$,
except if $|0_{2}\rangle$ is an eigen-vector of $P_{H}^{\left(2\right)}$,
i.e. if $H=\frac{1}{2}\alpha\left(q^{2}+p^{2}\right)$ is the Harmonic
oscillator. The geometrical meaning is that the linear symplectic
map $M\in\mbox{SL}\left(2,\mathbb{R}\right)$ does not preserves the
complex structure $J$ except if $M\in\mbox{U}\left(1\right)$ is
a rotation.
\item \textbf{Spectrum of the prequantum Harmonic oscillator:} With $\alpha=\beta=1$
and $\gamma=0$ in Eq.(\ref{eq:H_quad}), we obtain $H=\frac{1}{2}\left(q^{2}+p^{2}\right)$
. From Eq.(\ref{eq:PH_quadratic}), we observe that $P_{H}$ is the
sum of two {}``quantum Harmonic oscillators in 1::(-1) resonances'',
i.e. $P_{H}=\frac{1}{2}\left(\hat{Q}_{1}^{2}+\hat{P}_{1}^{2}\right)-\frac{1}{2}\left(\hat{Q}_{2}^{2}+\hat{P}_{2}^{2}\right)$.
We deduce that its spectrum $\sigma\left(P_{H}\right)$ is the set
of eigenvalues $\lambda_{n_{1},n_{2}}=\hbar\left(n_{1}+\frac{1}{2}\right)-\hbar\left(n_{2}+\frac{1}{2}\right)=\hbar\left(n_{1}-n_{2}\right)$,
with $n_{1},n_{2}\in\mathbb{N}$. So $\sigma\left(P_{H}\right)=\hbar\mathbb{Z}$,
with infinite multiplicity%
\footnote{More generally it could be interesting to compute the spectrum of
a prequantum operator if the classical Hamiltonian flow is integrable.%
}.
\end{itemize}
\begin{lem}
For any $v\in\mathbb{R}^{2}$, one trivially has $M\, T_{v}=T_{Mv}M$.
This conjugation relation persists at the prequantum level: \begin{equation}
\tilde{M}\,\tilde{T}_{v}=\tilde{T}_{Mv}\tilde{M}\label{eq:interviwining}\end{equation}
where $\tilde{T}_{v}$ is defined by Eq.(\ref{eq:PH_translation}),
and $\tilde{M}=\exp\left(-\frac{i}{\hbar}P_{H}\right)$ .
\end{lem}
\begin{proof}
For any point $x\in\mathbb{R}^{2}$ and $v\in\mathbb{R}^{2}$, the
linear relation $M\left(x+v\right)=M\left(x\right)+M\left(v\right)$
gives $MT_{v}=T_{Mv}M$. Consider the initial point $p=r_{x}\in L_{x}$
in the fiber over $x\in\mathbb{R}^{2}$. We want to compute the phase
$F$ obtained on the lifted path over the piece-wised closed path
$x=T_{v}^{-1}M^{-1}T_{Mv}M\left(x\right)$, defined by\begin{equation}
\tilde{T}_{v}^{-1}\tilde{M}^{-1}\tilde{T}_{Mv}\tilde{M}\left(p\right)=e^{-\frac{i}{\hbar}F}p\label{eq:phase_totale}\end{equation}
For a path generated by the quadratic Hamiltonian $H$, Eq.(\ref{eq:dF}),
Eq.(\ref{eq:F_quadratic}) gives that the phase is $F=0$. So only
the translations contribute to the phase. From Eq.(\ref{eq:F_translation})
and Eq.(\ref{eq:phase_totale}), we obtain:\begin{equation}
F=\frac{1}{2}\left(Mv\wedge Mx\right)-\frac{1}{2}v\wedge\left(x+v\right)=0\label{eq:F_si_H_quadratic}\end{equation}
using also the fact that $M$ preserves area.
\end{proof}

\section{\label{sec:Linear-cat-map}Linear cat map on the torus $\mathbb{T}^{2}$}

After the necessary presentation of prequantization on phase space
$\mathbb{R}^{2}$, we can now pass to the quotient $\mathbb{T}^{2}=\mathbb{R}^{2}/\mathbb{Z}^{2}$.
In this Section we recall the definition of the hyperbolic cat map
on the torus $\mathbb{T}^{2}=\mathbb{R}^{2}/\mathbb{Z}^{2}$ and present
its prequantization in the same way its quantization is usually obtained
(see e.g. \cite{berry-hannay-80},\cite{debievre96}, \cite{fred-steph-02}).

We start from a hyperbolic map \begin{equation}
M=\left(\begin{array}{cc}
A & B\\
C & D\end{array}\right)\in SL\left(2,\mathbb{Z}\right)\label{eq:M_T2}\end{equation}
 on $\mathbb{R}^{2}$, i.e. with integer coefficients such that $AD-BC=1$
and $\textrm{Tr}\left(M\right)=A+D>2$. A simple example is the {}``cat
map'' $M=\left(\begin{array}{cc}
2 & 1\\
1 & 1\end{array}\right)$\cite{arnold-avez}. For any $x\in\mathbb{R}^{2},n\in\mathbb{Z}^{2}$,
$M\left(x+n\right)=M\left(x\right)+M\left(n\right)\equiv M\left(x\right)\,\textrm{mod}1$
so $M$ induces a map on the torus $\mathbb{T}^{2}=\mathbb{R}^{2}/\mathbb{Z}^{2}$
also denoted by $M$, which is fully chaotic.

\subsection{Prequantum Hilbert space of the torus}

In this paragraph, we explicitly construct the prequantum Hilbert
space $\mathcal{\tilde{H}}_{N}$ associated to the torus phase space
$\mathbb{T}^{2}$, and the prequantum map $\tilde{M}\in\textrm{End}\left(\tilde{\mathcal{H}}_{N}\right)$
acting in it (respectively the quantum map $\hat{M}\in\textrm{End}\left(\mathcal{H}_{N}\right)$
acting in the quantum Hilbert space $\mathcal{H}_{N}$).

\subsubsection{Prequantum and Quantum Hilbert space for the torus $\mathbb{T}^{2}$
phase space}

The integer lattice $\mathbb{Z}^{2}\subset\mathbb{R}^{2}$ is generated
by the two vectors $\left(1,0\right)$ and $\left(0,1\right)$. We
consider the corresponding prequantum translation operators $\tilde{T}_{1}\defi\tilde{T}_{\left(1,0\right)}$
and $\tilde{T}_{2}\defi\tilde{T}_{\left(0,1\right)}$, defined by
Eq.(\ref{eq:translation_Ph}), which satisfy $\tilde{T}_{1}\tilde{T}_{2}=e^{-i/\hbar}\tilde{T}_{2}\tilde{T}_{1}$
as a result of Eq.(\ref{e:Heisenberg}). So for special values of
$\hbar$ given by:\[
\boxed{N=\frac{1}{2\pi\hbar}\in\mathbb{N}^{\star},}\]
one has the property $\left[\tilde{T}_{1},\tilde{T}_{2}\right]=0$.
We assume this relation from now on. 

We have seen in Eq.(\ref{eq:decomp_Tv}) that each operator has a
trivial action in the space $L^{2}\left(\mathbb{R}_{\left(2\right)}\right)$
entering the decomposition Eq.(\ref{eq:H1_H2}). So we will first
consider their action in the space $L^{2}\left(\mathbb{R}_{\left(1\right)}\right)$.
Let us define the space of {}`` periodic distributions''%
\footnote{We could have given a more general presentation with a decomposition
of $L^{2}\left(\mathbb{R}_{\left(1\right)}\right)$ into common eigenspaces
of the operators $\tilde{T}_{1}$,$\tilde{T}_{2}$: \begin{eqnarray*}
L^{2}\left(\mathbb{R}_{\left(1\right)}\right) & = & \int_{\left[0,2\pi\right]^{2}}^{\oplus}\mathcal{H}_{\left(1\right),N,\theta}\frac{d^{2}\theta}{\left(2\pi\right)^{2}},\\
\mathcal{H}_{\left(1\right),N,\theta} & \defi & \left\{ \psi_{\left(1\right)}\in\mathcal{S}'\left(\mathbb{R}_{\left(1\right)}\right)\,\textrm{such that }\tilde{T}_{1}\psi_{\left(1\right)}=e^{i\theta_{1}}\psi_{\left(1\right)},\,\tilde{T}_{2}\psi_{\left(1\right)}=e^{i\theta_{2}}\psi_{\left(1\right)}\right\} \end{eqnarray*}
with $\theta=\left(\theta_{1},\theta_{2}\right)\in\left[0,2\pi\right]^{2}$.
In this paper, we only consider the space $\mathcal{H}_{\left(1\right),N}=\mathcal{H}_{\left(1\right),N,\theta=0}$
which is sufficient for our purpose, and avoids more complicated notations.
See Section 3.2 in \cite{fred-steph-02}, where this more general
presentation is done.%
}: \begin{equation}
\mathcal{H}_{\left(1\right),N}\defi\left\{ \psi\in\mathcal{S}'\left(\mathbb{R}_{\left(1\right)}\right)\,\textrm{such that }\tilde{T}_{1}\psi=\psi,\,\tilde{T}_{2}\psi=\psi\right\} \label{eq:def_H1pre_N}\end{equation}

\paragraph{Characterization of the space $\mathcal{H}_{\left(1\right),N}$:}

\begin{lem}
$\textrm{dim}\mathcal{H}_{\left(1\right),N}=N$ . An explicit orthonormal
basis of $\mathcal{H}_{\left(1\right),N}$ is given by distributions
$\left(\varphi_{n}\right)_{n=0\ldots N-1}$ made of Dirac comb:\begin{equation}
\varphi_{n}\left(Q_{1}\right)=\frac{1}{\sqrt{N}}\sum_{k\in\mathbb{Z}}\delta\left(Q_{1}-\left(\frac{n}{N}+k\right)\right),\qquad n=0,\ldots,N-1\label{eq:phi_n_torus}\end{equation}

\end{lem}
\begin{proof}
First we observe that in the space $L^{2}\left(\mathbb{R}_{\left(1\right)}\right)$,
the operator $\tilde{T}_{1}=\tilde{T}_{\left(1,0\right)}=\exp\left(-\frac{i}{\hbar}\hat{P}_{1}\right)=\exp\left(-\frac{\partial}{\partial Q_{1}}\right)$
translates functions by one unit: $\left(\tilde{T}_{1}\psi\right)\left(Q_{1}\right)=\psi\left(Q_{1}-1\right)$,
and similarly the operator $\tilde{T}_{2}=\tilde{T}_{\left(0,1\right)}=\exp\left(-\frac{i}{\hbar}\left(-\hat{Q}_{1}\right)\right)$
translates the $\hbar-$Fourier Transform by one unit: $\left(\tilde{T}_{2}\hat{\psi}\right)\left(P_{1}\right)=\hat{\psi}\left(P_{1}-1\right)$,
with $\hat{\psi}\left(P_{1}\right)\defi\frac{1}{\sqrt{2\pi\hbar}}\int_{\mathbb{R}}\psi\left(Q_{1}\right)e^{-iP_{1}Q_{1}/\hbar}$.
So the space $\mathcal{H}_{\left(1\right),N}$ consists of distributions
$\psi\left(Q_{1}\right)$ which are periodic with period one, and
such that the Fourier transform is also periodic with period one.
As a result $\psi\left(Q_{1}\right)=\frac{1}{\sqrt{N}}\sum_{n\in\mathbb{Z}}\psi_{n}\,\delta\left(Q_{1}-nh\right)$
with $h=\frac{1}{N}=2\pi\hbar$, and with components $\psi_{n}\in\mathbb{C}$
which satisfy the periodicity relation $\psi_{n+N}=\psi_{n}$ . So
there are only $N$ independent components, and $\psi=\sum_{n=0}^{N-1}\psi_{n}\varphi_{n}$.
\end{proof}
Similarly to Eq.(\ref{eq:def_H1pre_N}), let us define the \textbf{prequantum
Hilbert space of the torus} by:\begin{equation}
\tilde{\mathcal{H}}_{N}\defi\left\{ \mbox{sections }s\in\Gamma^{\infty}\left(L\right)\,\textrm{such that }\tilde{T}_{1}s=s,\,\tilde{T}_{2}s=s,\qquad\int_{\left[0,1\right]^{2}}\left|s\left(x\right)\right|^{2}<\infty\right\} \label{eq:def_Hpre_N}\end{equation}

With the unitary isomorphism Eq.(\ref{eq:H1_H2}), and with Eq.(\ref{eq:def_H1pre_N}),
we can write%
\footnote{Note that this isomorphism gives an explicit orthonormal basis of
the prequantum Hilbert space $\tilde{\mathcal{H}}_{N}$ of $L^{2}$
sections of the Hermitian line bundle $L$ over $\mathbb{T}^{2}$,
which is not obvious a priori. Namely $\phi_{n,m}=\varphi_{n}\otimes\psi_{m}$
where $\varphi_{n},n=1\rightarrow N$, Eq.(\ref{eq:phi_n_torus}),
is an o.n. basis of $\mathcal{H}_{\left(1\right),N}$ and $\psi_{m},m\in\mathbb{N}$
is an orthonormal basis of $L^{2}\left(\mathbb{R}_{\left(2\right)}\right)$
(for example the eigenstates of the Harmonic oscillator given in Eq.(\ref{eq:op_a_ap})).
This basis has in fact a well known physical meaning: each space $\mathcal{H}_{\left(1\right),N}\otimes\left(\mathbb{C}\psi_{m}\right)$
is the eigenspace for the Hamiltonian of a free particle moving on
the torus $\mathbb{T}^{2}$ with a constant magnetic field $B=N\omega$.
The corresponding eigenvalues are called the Landau levels.%
}%
\footnote{The tensor product decomposition Eq.(\ref{eq:preq_space_torus}) which
is an important step in order to obtain Theorem \ref{thm:spectre}
can be considered as a simple (and surely well known) result of pure
representation theory of the Heisenberg group. More precisely, let
$H_{\mathbb{R}}$ be the Heisenberg group and $H_{\mathbb{Z}}$ be
the integral Heisenberg group. Then Eq.(\ref{eq:preq_space_torus})
concerns the decomposition of $L^{2}\left(H_{\mathbb{R}}\setminus H_{\mathbb{Z}}\right)$
under the action of $H_{\mathbb{R}}$ (whose Lie algebra is represented
in this paper by the operators $\hat{Q}_{2},\hat{P}_{2},Id$). %
}:\begin{equation}
\boxed{\tilde{\mathcal{H}}_{N}\equiv\mathcal{H}_{\left(1\right),N}\otimes L^{2}\left(\mathbb{R}_{\left(2\right)}\right).}\label{eq:preq_space_torus}\end{equation}

The definition Eq.(\ref{eq:def_Hpre_N}) is a space of sections of
$L\rightarrow\mathbb{R}^{2}$ periodic with respect to some action
of $\mathbb{Z}^{2}$. The space $\tilde{\mathcal{H}}_{N}$ can be
identified with the space of $L^{2}$ sections of a non trivial line
bundle $L\rightarrow\mathbb{T}^{2}$ over the torus, with Chern index
$N$. With respect to the trivialization $r$ the space $\tilde{\mathcal{H}}_{N}$
consists of quasi-periodic functions:\begin{equation}
\tilde{\mathcal{H}}_{N}\equiv\left\{ \psi\,\mbox{ s.t. }\psi\left(x+n\right)=\psi\left(x\right)e^{-i2\pi\frac{N}{2}n\wedge x}e^{-i2\pi\frac{N}{2}n_{1}n_{2}},\forall x\in\mathbb{R}^{2},\forall n\in\mathbb{Z}^{2}\textnormal{and }\int_{\left[0,1\right]^{2}}\left|\psi\left(x\right)\right|^{2}dx<\infty\right\} \label{eq:quasi-periodicity}\end{equation}

\begin{proof}
With Eq.(\ref{e:Heisenberg}) we have $\tilde{T}_{n}=\tilde{T}_{\left(n_{1},0\right)+\left(0,n_{2}\right)}=e^{i2\pi\frac{N}{2}n_{1}n_{2}}\tilde{T}_{\left(n_{1},0\right)}\tilde{T}_{\left(0,n_{2}\right)}=e^{i2\pi\frac{N}{2}n_{1}n_{2}}\tilde{T}_{1}^{n_{1}}\tilde{T}_{2}^{n_{2}}$.
Then with Eq.(\ref{eq:def_Hpre_N}), Eq.(\ref{eq:F_translation})
and Eq.(\ref{eq:Evolution_section}) $s=\psi r\in\tilde{\mathcal{H}}_{N}$
$\Leftrightarrow$ \{$\tilde{T}_{n}s=e^{i2\pi\frac{N}{2}n_{1}n_{2}}s$
for any $n\in\mathbb{Z}^{2}$\} $\Leftrightarrow\psi\left(x\right)e^{-i2\pi\frac{N}{2}n\wedge x}=e^{i2\pi\frac{N}{2}n_{1}n_{2}}\psi\left(x+n\right)$.
\end{proof}
In the same manner, let us define the \textbf{quantum Hilbert space
of the torus} by:

\begin{equation}
\mathcal{H}_{N}\defi\left\{ s\in\tilde{\mathcal{H}}_{N}\,\textrm{such that }s\mbox{ is anti-holomorphic}\right\} \label{eq:def_H_qu_N}\end{equation}

From Eq.(\ref{eq:H_quantum_L2R1}) we have:\[
\boxed{\mathcal{H}_{N}\equiv\mathcal{H}_{\left(1\right),N}\otimes\left(\mathbb{C}|0_{2}\rangle\right)\equiv\mathcal{H}_{\left(1\right),N}}\]

Note that there is a {}``perfect decoupling'' between the anti-holomorphic
condition which concerns the $L^{2}\left(\mathbb{R}_{\left(2\right)}\right)$
part of the decomposition Eq.(\ref{eq:H1_H2}), and the torus periodicity
which concerns the $L^{2}\left(\mathbb{R}_{\left(1\right)}\right)$
part.

\subsubsection{The prequantum cat map and the quantum cat map}

In order to obtain the prequantum map or quantum map corresponding
to $M:\mathbb{T}^{2}\rightarrow\mathbb{T}^{2}$ given in Eq.(\ref{eq:M_T2}),
we have first to describe $M$ as a Hamiltonian flow%
\footnote{The reason is essentially that a map itself has not all the information
necessary to define the prequantum or quantum map in a unique way.
In particular the {}``classical action'' of the trajectories are
not defined \emph{a priori.} If the map is obtained from a Poincaré
section or a stroboscopic section of a Hamiltonian flow, then there
is less arbitrariness to (pre)quantized it.%
}. The hyperbolic linear map $M:\mathbb{R}^{2}\rightarrow\mathbb{R}^{2}$,
$M\in\mbox{SL}\left(2,\mathbb{Z}\right)$, can be realized as a time
1 flow on $\mathbb{R}^{2}$ phase space generated by a \emph{hyperbolic
quadratic Hamiltonian function}:\begin{equation}
H\left(q,p\right)=\frac{1}{2}\alpha q^{2}+\frac{1}{2}\beta p^{2}+\gamma qp,\label{eq:H0}\end{equation}
From Hamiltonian equations $dq(t)/dt=\partial_{p}H=\gamma q+\beta p$,
$dp(t)/dt=-\partial_{q}H=-\alpha q-\gamma p$, we deduce that the
constants $\alpha,\beta,\gamma\in\mathbb{R}$ are obtained by solving
$M=\left(\begin{array}{cc}
A & B\\
C & D\end{array}\right)=\exp\left(\begin{array}{cc}
\gamma & \beta\\
-\alpha & -\gamma\end{array}\right)$. The \textbf{Lyapounov exponent} is given by $\lambda=\sqrt{\gamma^{2}-\alpha\beta}=\log\left(\frac{T+\sqrt{T^{2}-4}}{2}\right)$,
with $T=\mbox{Tr}\left(M\right)=A+D$, and gives the two eigenvalues
$e^{\pm\lambda}$ of $M$.

In Section \ref{sub:Case-of-a_Quadratic_H}, Eq.(\ref{eq:Mt_M1_M2}),
we have considered such quadratic Hamiltonian functions and obtained
that the prequantum map $\tilde{M}=\exp\left(-\frac{i}{\hbar}P_{H}\right)$
which is a unitary operator acting in $L^{2}\left(\mathbb{R}^{2}\right)\equiv L^{2}\left(\mathbb{R}_{\left(1\right)}\right)\otimes L^{2}\left(\mathbb{R}_{\left(2\right)}\right)$,
decomposes as $\tilde{M}=\tilde{M}_{\left(1\right)}\otimes\tilde{M}_{\left(2\right)}$. 

\vspace{0.5cm}\begin{center}\fbox{\parbox{15cm}{

\begin{lem}
If $N$ is even, the prequantum map $\tilde{M}$ in $L^{2}\left(\mathbb{R}^{2}\right)$
defines in a natural way unitary endomorphisms associated with the
torus phase space:\[
\tilde{M}_{\left(1\right),N}:\mathcal{H}_{\left(1\right),N}\rightarrow\mathcal{H}_{\left(1\right),N}\quad:\mbox{the quantum catmap}\]
\[
\tilde{M}_{N}\equiv\tilde{M}_{\left(1\right),N}\otimes\tilde{M}_{\left(2\right)}:\tilde{\mathcal{H}}_{N}\rightarrow\tilde{\mathcal{H}}_{N}\quad:\mbox{ the prequantum catmap}\]

\end{lem}
\selectlanguage{french}
}}\end{center}\vspace{0.5cm}

\selectlanguage{english}
\begin{proof}
Remind that the passage from the prequantum space $L^{2}\left(\mathbb{R}^{2}\right)\equiv L^{2}\left(\mathbb{R}_{\left(1\right)}\right)\otimes L^{2}\left(\mathbb{R}_{\left(2\right)}\right)$
to the torus prequantum space concerns only the $L^{2}\left(\mathbb{R}_{\left(1\right)}\right)$
part. Let us define a projector from the space $L^{2}\left(\mathbb{R}_{\left(1\right)}\right)$
onto the space $\tilde{\mathcal{H}}_{\left(1\right),N}$ by:\begin{equation}
\tilde{\mathcal{P}}_{\left(1\right)}\defi\sum_{\left(n_{1},n_{2}\right)\in\mathbb{Z}^{2}}\tilde{T}_{1}^{n_{1}}\tilde{T}_{2}^{n_{2}}=\sum_{\left(n_{1},n_{2}\right)\in\mathbb{Z}^{2}}\tilde{T}_{n},\label{eq:def_projector_P}\end{equation}
(we have used $\tilde{T}_{n}=\tilde{T}_{1}^{n_{1}}\tilde{T}_{2}^{n_{2}}$,
from Eq.(\ref{e:Heisenberg}), and the hypothesis that $N$ is even).
The domain of $\tilde{\mathcal{P}}_{\left(1\right)}$ consists of
fast decreasing sections. We extend $\tilde{\mathcal{P}}_{\left(1\right)}$
on the whole prequantum space $L^{2}\left(\mathbb{R}^{2}\right)\equiv L^{2}\left(\mathbb{R}_{\left(1\right)}\right)\otimes L^{2}\left(\mathbb{R}_{\left(2\right)}\right)$
by $\tilde{\mathcal{P}}=\tilde{\mathcal{P}}_{\left(1\right)}\otimes\hat{Id}_{\left(2\right)}$.
Using Eq.(\ref{eq:def_projector_P}) and Eq.(\ref{eq:interviwining}),
we have \[
\tilde{M}\tilde{\mathcal{P}}=\sum_{n\in\mathbb{Z}^{2}}\tilde{M}\tilde{T}_{n}=\sum_{n\in\mathbb{Z}^{2}}\tilde{T}_{Mn}\tilde{M}=\sum_{n\in\mathbb{Z}^{2}}\tilde{T}_{n}\tilde{M}=\tilde{\mathcal{P}}\tilde{M}\]
using that $M$ is one to one  on $\mathbb{Z}^{2}$. In particular
$\tilde{M}_{\left(1\right)}\tilde{\mathcal{P}}_{\left(1\right)}=\tilde{\mathcal{P}}_{\left(1\right)}\tilde{M}_{\left(1\right)}$.
This gives a commutative diagram:\begin{eqnarray*}
L^{2}\left(\mathbb{R}_{\left(1\right)}\right) & \underrightarrow{\,\,\tilde{M}_{\left(1\right)}\,\,\,} & L^{2}\left(\mathbb{R}_{\left(1\right)}\right)\\
\downarrow\tilde{\mathcal{P}}_{\left(1\right)} &  & \downarrow\tilde{\mathcal{P}}_{\left(1\right)}\\
\mathcal{H}_{\left(1\right),N} & \underrightarrow{\,\,\tilde{M}_{\left(1\right)}\,\,\,} & \mathcal{H}_{\left(1\right),N}\end{eqnarray*}
which means that $\tilde{M}_{\left(1\right)}$ induces a map denoted
$\tilde{M}_{\left(1\right),N}:\mathcal{H}_{\left(1\right),N}\rightarrow\mathcal{H}_{\left(1\right),N}$
(the quantum map), and similarly that $\tilde{M}$ induces a map denoted
by $\tilde{M}_{N}:\tilde{\mathcal{H}}_{N}\rightarrow\tilde{\mathcal{H}}_{N}$
(the prequantum map). The fact that $\tilde{M}_{\left(1\right),N}$
is the {}``usual'' quantum map is because its generator is obtained
by Weyl quantization in Eq.(\ref{eq:PH_1_quad}).
\end{proof}

\subsection{Prequantum resonances}

\subsubsection{Spectrum of the quantum map}

The spectrum of the quantum cat map, i.e. the unitary operator $\tilde{M}_{N,\left(1\right)}$
in the $N$ dimensional space $\mathcal{H}_{\left(1\right),N}$ is
well studied in the literature \cite{keating-91b}\cite{kurlberg-00}\cite{kurlberg-01}\cite{kurlberg-05}.
Let \begin{equation}
\tilde{M}_{N,\left(1\right)}|\psi_{\left(1\right),k}\rangle=e^{i\varphi_{k}}|\psi_{\left(1\right),k}\rangle,\quad k=1\rightarrow N\label{eq:eigen_values_M1}\end{equation}
 be the eigenvectors and eigenvalues of $\tilde{M}_{N,\left(1\right)}$.
(See figure \ref{fig:Comparison-of-spectra.} page \pageref{fig:Comparison-of-spectra.}). 

The prequantum map is the unitary map $\tilde{M}_{N}=\tilde{M}_{N,\left(1\right)}\otimes\tilde{M}_{\left(2\right)}$
acting in the infinite dimensional space $\mathcal{H}_{\left(1\right),N}\otimes L^{2}\left(\mathbb{R}_{\left(2\right)}\right)$.
The unitary operator $\tilde{M}_{\left(2\right)}=\exp\left(-\frac{i}{\hbar}P_{H}^{\left(2\right)}\right)$
is generated by $P_{H}^{\left(2\right)}=Op_{Weyl}\left(H_{\left(2\right)}\right)$,
with the \emph{hyperbolic} quadratic Hamiltonian $H_{\left(2\right)}$
given by Eq.(\ref{eq:H_2_quadrat}). $P_{H}^{\left(2\right)}$ has
a continuous spectrum with multiplicity two, therefore $\tilde{M}_{\left(2\right)}$
has a continuous spectrum on the unit circle. The spectrum of $\tilde{M}_{N}$
is then obtained by a product from the spectra of $\tilde{M}_{N,\left(1\right)}$
and $\tilde{M}_{\left(2\right)}$. The aim of this Section is to show
that $\tilde{M}_{\left(2\right)}$ and therefore $\tilde{M}_{N}$,
have nevertheless a well defined discrete spectrum of resonances.

\subsubsection{Normal form of the operator $\tilde{M}_{\left(2\right)}$}

We consider the operator $\tilde{M}_{\left(2\right)}=\exp\left(-\frac{i}{\hbar}P_{H}^{\left(2\right)}\right)$
with $P_{H}^{\left(2\right)}=Op_{Weyl}\left(H_{\left(2\right)}\right)=-\frac{1}{2}\alpha\hat{Q}_{2}^{2}-\frac{1}{2}\beta\hat{P}_{2}^{2}+\frac{\gamma}{2}\left(\hat{Q}_{2}\hat{P}_{2}+\hat{P}_{2}\hat{Q}_{2}\right)$
acting in the space $L^{2}\left(\mathbb{R}_{\left(2\right)}\right)$.
The classical symbol $H_{\left(2\right)}\left(q,p\right)=-\frac{1}{2}\alpha q^{2}-\frac{1}{2}\beta p^{2}+\gamma qp$
is a \emph{hyperbolic} quadratic function on $\mathbb{R}^{2}$. Therefore,
there exists a linear symplectic transformation $D\in SL\left(2,\mathbb{R}\right)$
which transforms $H_{\left(2\right)}$ into the hyperbolic normal
form:\[
N=H_{\left(2\right)}\circ D,\qquad N\left(q,p\right)=\lambda qp\]
with the Lyapounov exponent $\lambda=\sqrt{\gamma^{2}-\alpha\beta}$
(this last quantity is the unique symplectic invariant of the function
$H_{\left(2\right)}$).

At the operator level, there is a similar result: there is exists
a metaplectic operator (unitary operator in $L^{2}\left(\mathbb{R}_{\left(2\right)}\right)$),
given by $\hat{D}=\exp\left(-iOp_{Weyl}\left(d\right)/\hbar\right)$
(with $d\left(q,p\right)$ a quadratic form which generates $D$)
such that :\begin{equation}
\boxed{\hat{N}=\hat{D}P_{H}^{\left(2\right)}\hat{D}^{-1}=Op_{Weyl}\left(N\right)=\frac{\lambda}{2}\left(\hat{Q}_{2}\hat{P}_{2}+\hat{P}_{2}\hat{Q}_{2}\right)}\label{eq:def_N}\end{equation}

As a result, $\tilde{M}_{\left(2\right)}=\exp\left(-\frac{i}{\hbar}P_{H}^{\left(2\right)}\right)=\hat{D}^{-1}\exp\left(-\frac{i}{\hbar}\hat{N}\right)\hat{D}$
is conjugated to the normal form, so we can consider the operator
$\exp\left(-\frac{i}{\hbar}\hat{N}\right)$ or $\hat{N}$ itself,
which are simpler to handle.

\subsubsection{Quantum resonances of the quantum hyperbolic fixed point}

{}``Quantum resonances'' of $\hat{N}=Op_{Weyl}\left(\lambda qp\right)$
are well known. Note that with a canonical transform, $N\left(q,p\right)=\lambda qp$
is transformed to the inverted potential barrier: $H\left(x,\xi\right)=\frac{1}{2}\xi^{2}-\frac{1}{2}\lambda^{2}x^{2}$
. We recall here how to define and obtain these resonances by the
complex scaling method \cite{simon_87}. Consider first the classical
flow on $\left(\mathbb{R}^{2},dq\wedge dp\right)$ generated by the
hyperbolic Hamiltonian function $N\left(q,p\right)=\lambda qp$. The
point $\left(0,0\right)$ is a hyperbolic fixed point, with an unstable
direction $\left\{ p=0\right\} $, and a stable direction $\left\{ q=0\right\} $.
Let us introduce the quadratic {}``escape function'':\[
f_{\alpha}\left(q,p\right)=\frac{\alpha}{2}\left(p^{2}-q^{2}\right),\qquad\alpha>0\]
and define\[
\hat{f}_{\alpha}\defi Op_{Weyl}\left(f_{\alpha}\right),\quad\hat{A}_{\alpha}\defi\exp\left(\hat{f}_{\alpha}\right)\]
For $\left|\alpha\right|<\pi/2$, the domains $D_{A}\defi\mbox{dom}\left(\hat{A}_{\alpha}\right)$
and $C_{A}\defi\mbox{dom}\left(\hat{A}_{\alpha}^{-1}\right)$ are
dense in $L^{2}\left(\mathbb{R}_{\left(2\right)}\right)$. One can
explicitly check that they contain Gaussian wave functions. The choice
of the escape function $f_{\alpha}$ is related to the property that
it decreases along the flow of $N$: let $X_{N}=\lambda\left(q\frac{\partial}{\partial q}-p\frac{\partial}{\partial p}\right)$
be the Hamiltonian vector field associated to $N$, then $X_{N}\left(f_{\alpha}\right)=-\alpha\lambda\left(q^{2}+p^{2}\right)<0$
if $q,p\neq0$.

\begin{lem}
For $\left|\alpha\right|<\pi/2$, let $\hat{K}_{\alpha}\defi\frac{i}{\hbar}\hat{A}_{\alpha}\hat{N}\hat{A}_{\alpha}^{-1}$.
Then $\hat{K}_{\alpha}$ is defined on a dense domain in $L^{2}\left(\mathbb{R}_{\left(2\right)}\right)$,
and \[
\hat{K}_{\alpha}=\frac{1}{\hbar}\lambda\sin\left(2\alpha\right)Op_{Weyl}\left(\frac{1}{2}\left(q^{2}+p^{2}\right)\right)+\frac{i}{\hbar}\lambda\cos\left(2\alpha\right)Op_{Weyl}\left(qp\right)\]
In particular for $\alpha=\frac{\pi}{4}$, \begin{equation}
\hat{K}_{\pi/4}=\lambda\frac{1}{2\hbar}\left(\hat{q}^{2}+\hat{p}^{2}\right)\label{eq:K_OH}\end{equation}
 is the quantum Harmonic oscillator with discrete spectrum $\lambda_{n}=\lambda\left(n+\frac{1}{2}\right)$,
$n\in\mathbb{N}$.
\end{lem}
We keep now the simple choice $\alpha=\pi/4$, and write $\hat{K}\defi\hat{K}_{\pi/4}$,
$\hat{A}\defi\hat{A}_{\pi/4}$.

\begin{proof}
The proof requires some standard calculation with the complexified
metaplectic group, whose Lie algebra $sp\left(2,\mathbb{R}\right)^{\mathbb{C}}=sl\left(2,\mathbb{C}\right)$
is generated by the three operators $Op_{Weyl}\left(qp\right)$, $Op_{Weyl}\left(\frac{1}{2}\left(p^{2}-q^{2}\right)\right)$,
$Op_{Weyl}\left(\frac{1}{2}\left(p^{2}+q^{2}\right)\right)$, see
\cite{folland-88} chapter 4, or \cite{ec1} p. 896.
\end{proof}
\begin{cor}
Let $\hat{B}\defi\hat{A}\hat{D}$. By a non (unitary) conjugation,
$\tilde{M}_{\left(2\right)}=\exp\left(-\frac{i}{\hbar}P_{H}^{\left(2\right)}\right)$
is transformed on a dense domain, into a Trace class operator:\begin{equation}
\hat{R}\defi\hat{B}\tilde{M}_{\left(2\right)}\hat{B}^{-1}=\exp\left(-\hat{K}\right)\label{eq:conjugate_M2}\end{equation}
with eigenvalues\[
\exp\left(-\lambda_{n}\right),\qquad\lambda_{n}\defi\lambda\left(n+\frac{1}{2}\right),\qquad n\in\mathbb{N}\]

\end{cor}

\paragraph{Remarks: }

\begin{itemize}
\item We would have obtain the same result with any choice of $0<\alpha<\pi/2$. 
\item Because $\hat{R}$ is defined on a dense domain, and is a bounded
operator, it extends in a unique way to $L^{2}\left(\mathbb{R}_{\left(2\right)}\right)$.
The eigenvalues $\exp\left(-\lambda_{n}\right)$ are called the {}``quantum
resonances'' of the unitary operator $\tilde{M}_{\left(2\right)}$.
The meaning of the operator $\hat{R}$ and its eigenvalues, appears
in the study of the decay of time-correlation functions. If $\varphi\in D_{C}=\mbox{dom}\left(\hat{B}\right)$,
and $\phi\in C_{C}=\mbox{dom}\left(\hat{B}^{-1}\right)$ are suitable
functions, then $C_{\phi,\varphi}\left(t\right)\defi\langle\phi|\tilde{M}_{\left(2\right)}^{t}|\varphi\rangle$,
$t\in\mathbb{N}$, can be expressed using $\hat{R}$ as\[
C_{\phi,\varphi}\left(t\right)=\left(\langle\phi|\hat{B}^{-1}\right)\hat{R}^{t}\left(\hat{B}|\varphi\rangle\right)\]
Then, the spectrum of $\hat{R}$ gives the explicit exponential decay
of the time-correlation function $C_{\phi,\varphi}\left(t\right)$.
The decay comes from the simple fact that there is an unstable fixed
point at the origin, and therefore the wave function $\varphi_{t}=\tilde{M}_{\left(2\right)}^{t}\varphi$
spreads along the unstable direction. This is very general in physics
and mathematics \cite{zworski_resonances_99}.
\end{itemize}

\subsubsection{\label{sub:proof}Resonances of the prequantum operator}

The conjugation operator $\hat{B}=\hat{A}\hat{D}$ has been defined
on $L^{2}\left(\mathbb{R}_{\left(2\right)}\right)$ and can be extend
to the prequantum space $\tilde{\mathcal{H}}_{N}\equiv\mathcal{H}_{\left(1\right),N}\otimes L^{2}\left(\mathbb{R}_{\left(2\right)}\right)$
by $\tilde{B}\defi\hat{\mbox{Id}}_{\left(1\right)}\otimes\hat{B}$.
We use it to conjugate the prequantum map $\tilde{M}_{N}=\tilde{M}_{\left(1\right),N}\otimes\tilde{M}_{\left(2\right)}$
and deduce from Eq.(\ref{eq:eigen_values_M1}) and Eq.(\ref{eq:conjugate_M2}): 

\begin{thm}
The conjugated operator \begin{equation}
\tilde{R}\defi\tilde{B}\tilde{M}_{N}\tilde{B}^{-1}=\tilde{M}_{N,\left(1\right)}\otimes\hat{R}\label{eq:M_conjugue_op_trace}\end{equation}
 is a Trace class operator in the prequantum space $\tilde{\mathcal{H}}_{N}\equiv\mathcal{H}_{\left(1\right),N}\otimes L^{2}\left(\mathbb{R}_{\left(2\right)}\right)$,
with eigenvalues:\[
r_{n,k}\defi\exp\left(i\varphi_{k}-\lambda_{n}\right),\qquad\lambda_{n}\defi\lambda\left(n+\frac{1}{2}\right),\,\, n\in\mathbb{N},\qquad\varphi_{k}\in\left[0,2\pi\right],\, k\in\left[1,N\right]\]

\end{thm}
The eigenvalues $r_{n,k}$ are called the \textbf{resonances of the
prequantum} \textbf{map}. This gives Theorem \ref{thm:spectre} page
\pageref{thm:spectre}, the main result of this paper.

\subsection{\label{sub:Relation-between-prequantum}Relation between prequantum
time-correlation functions and quantum evolution of wave functions}

Let $\varphi,\phi\in\tilde{\mathcal{H}}_{N}$ be prequantum wave functions,
who belong respectively to the domains of $\tilde{B}$ and $\tilde{B}^{-1}$.
Let us define $\tilde{\phi}=\hat{\Pi}\tilde{B}^{-1}\phi$, $\tilde{\varphi}=\hat{\Pi}\tilde{B}\varphi$
, where $\hat{\Pi}=\hat{I}_{1}\otimes|0_{2}\rangle\langle0_{2}|:\tilde{\mathcal{H}}_{N}\rightarrow\mathcal{H}_{\left(1\right),N}$
is the orthogonal Toeplitz projector. Then $\langle\phi|\tilde{M}_{N}^{t}|\varphi\rangle=\langle\phi|\tilde{M}_{\left(1\right),N}^{t}\otimes\tilde{M}_{\left(2\right)}^{t}|\varphi\rangle$,
but $\tilde{M}_{\left(2\right)}^{t}=\tilde{B}^{-1}\tilde{R}^{t}\tilde{B}$
and $\tilde{R}^{t}=\sum_{n_{2}\in\mathbb{N}}|n_{2}\rangle\langle n_{2}|\exp\left(-\lambda\left(n_{2}+\frac{1}{2}\right)t\right)$.
We deduce that $\langle\phi|\tilde{M}_{N}^{t}|\varphi\rangle=\langle\phi|\tilde{M}_{\left(1\right),N}^{t}\otimes\left(\tilde{B}^{-1}|0_{2}\rangle\langle0_{2}|\tilde{B}\right)|\varphi\rangle e^{-\lambda t/2}\left(1+\mathcal{O}\left(e^{-\lambda t}\right)\right)$,
hence\[
\langle\phi|\tilde{M}_{N}^{t}|\varphi\rangle=\langle\tilde{\phi}|\tilde{M}_{\left(1\right),N}^{t}|\tilde{\varphi}\rangle e^{-\lambda t/2}\left(1+\mathcal{O}\left(e^{-\lambda t}\right)\right)\]
This gives Proposition \ref{pro:quantum_correl} page \pageref{pro:quantum_correl}.
Let us remark that $|0_{2}\rangle$ does not belong to the domains
of $\tilde{B}$ or $\tilde{B}^{-1}$, but $\tilde{B}|0_{2}\rangle$
can be interpreted as a distribution, so $\langle0_{2}|\tilde{B}|\varphi\rangle$
makes sense even if $\varphi$ does not belong to the domain of $\tilde{B}$.

\subsection{Proof of the trace formula\label{sub:Proof-of-the_trace_formula}}

We prove here Proposition \ref{pro:Trace_formula_Rt} page \pageref{pro:Trace_formula_Rt}.
We just follow the calculation of Eq.(\ref{eq:calcul_trace}), but
with a suitable regularization, and show that it gives $\mbox{Tr}\left(\tilde{R}^{t}\right)$.
We follows a similar calculation which has been made in \cite{fred-RP-06}.
This proof does not used crucially the hypothesis that $M$ is a linear
map, so it could work for non linear prequantum hyperbolic map as
well.

Let us introduce a cutoff operator in space $L^{2}\left(\mathbb{R}_{\left(2\right)}\right)$
defined in Eq. (\ref{eq:H1_H2}):\[
P_{\nu}\defi\exp\left(-\nu\frac{1}{2}\left(\hat{P}_{2}^{2}+\hat{Q}_{2}^{2}\right)\right),\qquad\nu>0\]
This operator is diagonal in the basis $|n_{2}\rangle$ of the Harmonic
oscillator, and truncates high values of $n_{2}$. We choose here
a metaplectic operator for future convenience. The operator $P_{\nu}$
is Trace Class, and converges strongly towards identity for $\nu\rightarrow0$.
Consequently, $\langle Q_{2}'|P_{\nu}|Q_{2}\rangle\rightarrow\delta\left(Q_{2}'-Q_{2}\right)$
for $\nu\rightarrow0$ and uniformly with respect to $Q_{2}\in K\subset\mathbb{R}_{\left(2\right)}$
in a compact set. We extend this operator in $\tilde{\mathcal{H}}_{N}=\mathcal{H}_{\left(1\right),N}\otimes L^{2}\left(\mathbb{R}_{\left(2\right)}\right)$
by $Id_{\left(1\right)}\otimes P_{\nu}$ and denote it again $P_{\nu}$.
Using Eq.(\ref{eq:formula_qp_Q1Q2}) it is possible to show that $\langle x'|P_{\nu}|x\rangle\underset{\nu\rightarrow0}{\longrightarrow}\delta\left(x-x'\right)$
uniformly with respect to $x\in K\subset\mathbb{R}^{2}$ in a compact
set.

\begin{lem}
\label{lem:trace1}For any $t>0$, $\nu>0$, $\left(\tilde{M}_{N}^{t}P_{\nu}\right)$
is a Trace Class operator in $\tilde{\mathcal{H}}_{N}$ and\[
\mbox{Tr}\left(\tilde{M}_{N}^{t}P_{\nu}\right)\underset{\nu\rightarrow0}{\longrightarrow}\sum_{x\equiv M^{t}x\,\left[1\right]}\frac{1}{\left|\textrm{det}\left(1-M^{t}\right)\right|}e^{iA_{x,t}/\hbar}\]
where the sum is over points $x\in[0,1[^{2}$ such that $M^{t}x=x+n$,
with $n\in\mathbb{Z}^{2}$, i.e. periodic points on $\mathbb{T}^{2}$.
$A_{x,t}=\frac{1}{2}n\wedge x$ is the {}``classical action'' of
the periodic point $x$.
\end{lem}
\begin{proof}
First $\tilde{M}_{N}^{t}P_{\nu}$ is Trace class because it is a product
of a unitary and Trace class operator. Using Eq.(\ref{eq:Evolution_section})
for the prequantum evolution and Dirac notations, we write\[
\left(\tilde{M}^{t}\psi\right)\left(x\right)=\langle x|\tilde{M}^{t}|\psi\rangle=\psi\left(M^{-t}x\right)e^{-iF_{M^{-t}x,t}/\hbar}=\psi\left(M^{-t}x\right)\]
because since $M$ is linear, we have shown in Eq.(\ref{eq:F_quadratic})
that the phase is $F_{x,t}=0$. Then with $|\psi_{x,\nu}\rangle\defi P_{\nu}|x\rangle$,
the operator $\tilde{\mathcal{P}}$ defined in Eq.(\ref{eq:def_projector_P}),
and using $\tilde{T}_{n}|x\rangle=e^{-i\frac{1}{2\hbar}n\wedge x}|x+n\rangle$,
\begin{align*}
\mbox{Tr}\left(\tilde{M}_{N}^{t}P_{\nu}\right) & =\int_{]0,1[^{2}}\langle x|\tilde{\mathcal{P}}\tilde{M}^{t}P_{\nu}|x\rangle dx=\int_{]0,1[^{2}}\langle x|\tilde{\mathcal{P}}\tilde{M}^{t}|\psi_{x,\nu}\rangle dx\\
 & =\sum_{n\in\mathbb{Z}^{2}}\int_{]0,1[^{2}}e^{i\frac{1}{2\hbar}n\wedge x}\langle x+n|\tilde{M}^{t}|\psi_{x,\nu}\rangle dx\\
 & =\sum_{n\in\mathbb{Z}^{2}}\int_{]0,1[^{2}}e^{i\frac{1}{2\hbar}n\wedge x}\psi_{x,\nu}\left(M^{-t}\left(x+n\right)\right)dx\\
\\\end{align*}
We have seen that $\psi_{x,\nu}\left(x'\right)=\langle x'|P_{\nu}|x\rangle\underset{\nu\rightarrow0}{\longrightarrow}\delta\left(x-x'\right)$
uniformly with respect to $x$ in a compact set, so \begin{align*}
\mbox{Tr}\left(\tilde{M}_{N}^{t}P_{\nu}\right)\underset{\nu\rightarrow0}{\longrightarrow}\sum_{n\in\mathbb{Z}^{2}}\int_{]0,1[^{2}}\delta\left(x-M^{-t}\left(x+n\right)\right)e^{i\frac{1}{2\hbar}n\wedge x}dx\\
=\sum_{x\equiv M^{t}x\,\left[1\right]}\frac{1}{\left|\textrm{det}\left(1-M^{t}\right)\right|}e^{i\frac{1}{2\hbar}n\wedge x}\end{align*}
We have used a change of variable $x\rightarrow y=x-M^{-t}x-M^{-t}n$,
and where a periodic point $x\in]0,1[^{2}$ is specified by $M^{t}x=x+n$,
$n\in\mathbb{Z}^{2}$.
\end{proof}
\begin{lem}
\label{lem:trace2}$\mbox{Tr}\left(\tilde{M}^{t}P_{\nu}\right)\underset{\nu\rightarrow0}{\longrightarrow}\mbox{Tr}\left(\tilde{R}^{t}\right)$.
\end{lem}
\begin{proof}
We have $\tilde{M}^{t}P_{\nu}=\tilde{B}^{-1}\tilde{R}^{t}\tilde{B}P_{\nu}$,
then $\mbox{Tr}\left(\tilde{M}^{t}P_{\nu}\right)=\mbox{Tr}\left(\tilde{R}^{t}\tilde{B}P_{\nu}\tilde{B}^{-1}\right)$.
This involves a product of metaplectic operators, and using a representation
of $SL\left(2,\mathbb{C}\right)$, we explicitly check that $\tilde{R}^{t}\tilde{B}P_{\nu}\tilde{B}^{-1}$
converges towards $\tilde{R}^{t}$ as $\nu\rightarrow0$. (Notice
that for non linear maps, this arguments would have failed, and the
proof would have been longer).
\end{proof}
With Lemma \ref{lem:trace1} and Lemma \ref{lem:trace2} taken together,
we conclude the proof of Proposition \ref{pro:Trace_formula_Rt}.

\section{Conclusion}

In this paper we have defined the prequantum map associated to a linear
hyperbolic map on the torus $\mathbb{T}^{2}$, and shown that it has
well defined resonances. These resonances form a discrete spectrum
and can be explicitely expressed with the eigenvalues of the unitary
quantum map. In Section \ref{sec:Statement-of-the}, we have discussed
the interpretation of this spectrum of resonances in terms of decay
of time correlation functions, and compared them with the matrix elements
of the quantum map after time $t$. We have also compared the trace
formula for the quantum propagator and for the prequantum one (the
sum over its resonances) after a large time.

We would like first to make a general remark on prequantum dynamics.
Prequantization is well known since many years, and it is known to
be a very beautiful theory from a geometrical point of view. Many
works have study the geometrical aspects, and show how to define prequantization
in very general cases, for example Hodge manifolds. From a mathematical
perspective in dynamical systems, prequantization is directly defined
from the Hamiltonian flow, so that it is natural to investigate its
properties, for example, its spectrum. Nevertheless, it seems that
very few work have already investigate its dynamical properties and
its spectrum. This paper goes in this direction, and we would like
to emphasize that prequantum spectrum is not only interesting by itself,
but may rather be a useful approach for semi-classical analysis, especially
for quantum hyperbolic dynamics, i.e. {}``quantum chaos''. 

It is natural to ask if such results have been investigated for the
geodesic flow on negative curvature manifold. In fact, in the case
of cotangent phase spaces, the prequantum bundle is trivial, and the
prequantum operator can be expressed as a classical transfer operator
with a suitable weighted function. Such an operator is well studied
and it is known that the spectrum of classical resonances for the
geodesic flow on constant negative curvature is related to the spectrum
of the Laplacian which plays the role of the quantum operator (the
relation can be obtained using the Selberg zeta function \cite{venkov_90},
or by group theory approach in \cite{ratner_87}).

Some interesting questions arrive naturally in the framework of prequantum
chaos, similar to questions which exist in quantum chaos, namely concerning
the {}``semi-classical limit'' $N=1/\left(2\pi\hbar\right)\rightarrow\infty$,
where the curvature of the prequantum bundle goes to infinity. If
properly defined, one could investigate the problem of {}``prequantum
ergodicity'' or {}``unique prequantum ergodicity''. For example,
in \cite{fred-steph-02}, the existence of scarred quantum eigenfunctions
has been obtained, i.e. non equidistributed eigen-functions over the
torus in the limit $N\rightarrow\infty$. Because of the explicit
relation between quantum eigenfunctions and prequantum resonances
we have obtained, this could lead to {}``prequantum scarred distributions''
(but this needs some correct definition). Let us remark that the Ehrenfest
time $t_{E}\defi\frac{1}{\lambda}\log N$ is known to play an important
role as a characteristic time scale in quantum chaos \cite{fred_ihp_05}.
Its usual interpretation is the time after which a detail of the size
of $\hbar$, i.e. the minimum size in phase space allowed by the quantum
uncertainty principle, called the Planck cell, is exponentially amplified
towards finite size: $\hbar e^{\lambda t_{E}}\simeq1$. In prequantum
dynamics, there is no more uncertainty principle because the dynamics
evolves smooth sections over phase space. But the prequantum bundle
has a curvature $\Theta=\frac{i}{\hbar}\omega$, and there is still
the notion of Planck cell on the torus as the elementary surface over
which the curvature integral is one. Therefore the Ehrenfest time
may still play an important role for prequantum dynamics, at least
in the semi-classical limit $N\rightarrow\infty$.

\paragraph{Perspectives in the non linear case:}

For a linear map $M$, we have shown that there is an exact correspondence
between the spectrum of prequantum resonances and the quantum spectrum.
In a future work we plan to study non linear prequantum hyperbolic
map on the torus, and expect to obtain similar results%
\footnote{Let us remark that structural stability theorem guaranties that the
prequantum dynamics is still hyperbolic.%
} (with possibly introducing some weight function $\varphi=\lambda/2$
in the transfer operator, where $\lambda$ is the local expanding
rate). We expect then that there still exists an exact prequantum
trace formula for $\textrm{Tr}\left(\tilde{R}_{\varphi}^{t}\right)$
in terms of periodic orbits, similar to Eq.(\ref{eq:Trace_M_preq}).
We hope to be able to compare the prequantum operator $\tilde{R}_{\varphi}$
with the quantum operator $\hat{M}$, and possibly their spectra as
we did in Eq.(\ref{eq:spectre_r_nk}), at least in the limit $N\rightarrow\infty$,
and then deduce validity of the semi-classical Gutzwiller trace formula
and other semi-classical formula for long times. Some interesting
questions would appear then, as: does the random matrix theory applies
for the outlying prequantum spectra? 

\bibliographystyle{plain}
\bibliography{/home/faure/articles/articles}

\end{document}